\documentstyle[12pt,aaspp4]{article}
\newcommand{\kms}{km s$^{-1}$}
\newcommand{\zabs}{$z_{\rm abs}$}
\newcommand{\lya}{Ly$\alpha$\ }
\newcommand{\nav}{$N_{\rm a}(v)$}
\slugcomment{\footnotesize To appear in {\bf The Astrophysical Journal}, Oct. 10, 2000 (vol. 542)}
\lefthead{TRIPP \& SAVAGE}
\righthead{LOW REDSHIFT O VI}
\begin{document}
\title{\ion{O}{6} and Multicomponent \ion{H}{1} Absorption Associated with
a Galaxy Group in the Direction of PG0953+415: Physical Conditions and 
Baryonic Content\footnote{Based on observations 
obtained with the WIYN Observatory, which is a joint facility of the 
University of Wisconsin, Indiana University, Yale University, and the National 
Optical Astronomy Observatories.}$^{,}$\footnote{Based on observations 
with the NASA/ESA {\it Hubble Space Telescope}, obtained at the Space 
Telescope Science Institute, which is operated by the Association of 
Universities for Research in Astronomy, Inc., under NASA contract NAS 
5-2555.}}

\author{Todd M. Tripp\altaffilmark{3} and Blair D. 
Savage\altaffilmark{4}}

\altaffiltext{3}{Princeton University Observatory, 
Peyton Hall, Princeton, NJ 08544, 
Electronic mail: tripp@astro.princeton.edu}

\altaffiltext{4}{Department of Astronomy, University of Wisconsin 
- Madison, 475 N. Charter St., Madison, WI 53706 - 1582, 
Electronic mail: savage@astro.wisc.edu}

\begin{abstract}
We report the discovery of an \ion{O}{6} absorption system at \zabs\ = 0.14232 
in a high resolution FUV spectrum of PG 0953+415 obtained with the Space 
Telescope Imaging Spectrograph (STIS). Both lines of the \ion{O}{6} $\lambda 
\lambda$ 1032, 1038 doublet and multicomponent \ion{H}{1} \lya absorption are 
detected, but the \ion{N}{5} doublet and the strong lines of \ion{C}{2} and 
\ion{Si}{3} are not apparent. We examine the ionization mechanism of the 
\ion{O}{6} absorber and find that while theoretical considerations favor 
collisional ionization, it is difficult to observationally rule out  
photoionization. If the absorber is collisionally ionized, it may not be in 
equilibrium due to the rapid cooling of gas in the appropriate temperature 
range. Non-equilibrium collisionally ionized models are shown to be consistent 
with the observations. A WIYN survey of galaxy redshifts near the sight line 
has revealed a galaxy at a projected distance of 395 kpc separated by 
$\sim$130 \kms\ from this absorber, and three additional galaxies are found 
within $\lesssim$ 130 \kms\ of this redshift with projected separations 
ranging from 1.0 Mpc to 3.0 Mpc. All of these galaxies are luminous 
(0.6 -- 4.0 $L*$), and two of them show the [\ion{O}{2}] $\lambda$3727 
emission line indicative of active star formation. The galaxies with 
[\ion{O}{2}] emission are probably normal spirals. Combining the STIS 
observations of PG0953+415 with previous high signal-to-noise observations 
of H1821+643 with the Goddard High Resolution Spectrograph (GHRS), 
we find two \ion{O}{6} systems with $W_{\rm r} >$ 60 m\AA\ 
and $z <$ 0.3 over a total redshift path $\Delta z$ of only 0.10. Both of 
these QSOs were originally observed to study the low $z$ \lya lines and 
should not be biased in favor of \ion{O}{6} detection.  If these sight lines 
are representative, they imply a large number of \ion{O}{6} absorbers per 
unit redshift, $dN/dz \sim$ 20. The corresponding value of $dN/dz$ for low 
$z$ \lya lines with $W_{\rm r} >$ 50 m\AA\ is $102 \pm 16$.  We use this 
sample to obtain a first estimate of the cosmological mass density of the 
\ion{O}{6} systems at $z \approx$ 0.  If further observations confirm the 
large $dN/dz$ derived for the \ion{O}{6} systems, then these absorbers trace 
a significant reservoir of baryonic matter at low redshift.
\end{abstract}

\keywords{cosmology: observations --- galaxies: halos --- 
intergalactic medium --- quasars: absorption lines --- 
quasars: individual (PG0953+415)}

\section{Introduction}

Hydrodynamic simulations of cosmological structure growth predict that 
when the initial density perturbations collapse, gas should be 
shock-heated to temperatures of $10^{5}$ -- $10^{7} \ ^{\circ}$K 
(Ostriker \& Cen\markcite{oc96} 1996; Dav\'{e} et al.\markcite{dave99} 
1999; Cen \& Ostriker\markcite{co99a} 1999a). The fraction of the gas 
which has been heated to these temperatures increases with decreasing 
redshift, and at the present epoch, the model of Cen \& 
Ostriker\markcite{co99a} (1999a) predicts that 47 \% of the baryons (by 
mass) are in this shock-heated phase, hereafter referred to as warm/hot 
gas (to distinguish it from the hotter gas in rich galaxy clusters 
which are readily detected X-ray sources). This warm/hot gas prediction has 
not been adequately tested by observations because the soft X-rays emitted by 
gas at these temperatures are difficult to detect with current instrumentation, 
especially at lower temperatures where corrections for foreground 
absorption and emission are complicated. However, it may 
be possible to detect gas in the lower half of this temperature range 
via absorption lines of species such as \ion{O}{6}, \ion{Ne}{8}, or 
\ion{Mg}{10} in the spectrum of a background QSO (Verner, Tytler, \& 
Barthel\markcite{vern94} 1994). It is important to search for this warm/hot 
gas as part of the census of matter in the universe and because it could
affect the formation and evolution of galaxies and galaxy groups and clusters
(e.g., Blanton et al.\markcite{blan99} 2000).

There are some indications that warm/hot gas is present in some galaxy 
groups. For example, Mulchaey et al.\markcite{mul96} 
(1996) have noted that ROSAT observations show that poor galaxy groups 
which are rich in elliptical galaxies tend to exhibit X-ray emission (E $>$ 
0.5 keV) while spiral-rich groups do not. They suggest that spiral-rich groups 
contain cooler ($\lesssim 4\times 10^{6}$ K) intragroup gas which is 
not easily detected in X-rays, and they predict that the intragroup medium 
of spiral-rich groups will produce absorption lines of \ion{O}{6}, but not 
\ion{C}{4} or \ion{N}{5} because their column densities are too low at 
$T > 5 \times 10^{5} \ ^{\circ}$K. Interestingly, Savage, Tripp, \& 
Lu\markcite{stl98} (1998) have recently identified a QSO absorber in 
the spectrum of H1821+643 ($z_{\rm QSO}$ = 0.297) which fits this 
description: \ion{O}{6} absorption lines with two nearby spiral 
galaxies but no accompanying \ion{C}{4} or \ion{N}{5} lines. However, 
the absorption could be due to the halo of the closer spiral 
galaxy (which is at a projected distance of 105 $h_{75}^{-1}$ kpc) 
rather than the intragroup medium, and Savage et al.\markcite{stl98} 
show that the \ion{O}{6} absorption could plausibly arise in very low density 
photoionized gas. The \ion{O}{6} doublet has been 
identified in several other intervening absorption systems\footnote{Strong 
O VI absorption lines have also been detected in ``associated''
absorption line systems with \zabs\ $\approx \ z_{\rm QSO}$ (e.g., Papovich 
et al.\markcite{pap} 2000 and references therein). These are a rather different
class of absorber which are often known to be very close to the QSO (Hamann 
\& Ferland\markcite{hf99} 1999). In this paper, we have focused our analysis
and discussion on the intervening systems.} both at moderate 
redshifts (e.g., Bergeron et al.\markcite{berg94} 1994; Burles \& 
Tytler\markcite{bt96} 1996; Jannuzi et al.\markcite{jan98} 1998; 
Lopez et al.\markcite{lop99} 1999; Churchill \& Charlton\markcite{cc99} 1999) 
and at high redshifts (Kirkman \& Tytler\markcite{kt97} \markcite{kt99} 
1997,1999). Composite spectra and statistical techniques have also been 
used to show that \ion{O}{6} absorption is present at high $z$ (Lu \& 
Savage\markcite{ls93} 1993; Dav\'{e} et al.\markcite{dave98} 1998). In most 
cases it has been difficult to pin down the ionization mechanism definitively, 
partly due to the low resolution of the observations made with 
first-generation {\it Hubble Space Telescope (HST)} spectrographs, but in 
several cases there is evidence that the \ion{O}{6} systems occur in multiphase 
absorbing media. An important step in this approach to the search for 
warm/hot gas is to determine whether the \ion{O}{6} absorbers 
trace collisionally ionized gas or photoionized gas.

As part of the program described by Tripp, Lu, \& Savage\markcite{tls98} 
(1998) to study low $z$ \lya absorption line systems, we have recently 
observed the low redshift QSO PG0953+415 ($z_{\rm QSO}$ = 0.239) with the 
E140M echelle mode of the Space Telescope Imaging Spectrograph (STIS). 
This high resolution FUV spectrum has revealed another highly ionized 
\ion{O}{6} absorber associated with a group of spiral galaxies, and in 
this paper we present our analysis of this particular absorbing system. 
In \S 2 we review the observations and data reductions including measurements 
of galaxy redshifts with the WIYN telescope. We present in \S 3 the 
absorption line measurements. We constrain the temperature of the gas 
and examine its ionization in \S 4, and we discuss the implications of 
the observations in \S 5. Throughout this paper we assume $H_{0} = 75 
h_{75}$ \kms\ Mpc$^{-1}$ and $q_{0}$ = 0.0. Also, all wavelengths and 
redshifts reported here are heliocentric, but in this direction heliocentric
and LSR velocities are nearly identical.\footnote{Assuming the Sun is 
moving in the direction $l = 53^{\circ}, b = 25^{\circ}$ at 16.5 \kms\ 
(Mihalas \& Binney\markcite{mb81} 1981), $v_{\rm LSR} = v_{\rm helio} -
0.1$ \kms .}

\section{Observations}

\subsection{STIS Spectroscopy of PG 0953+415}

PG 0953+415 was observed with STIS on 1998 December 4 and 1998 December 
11 resulting in a total integration time of 24478 seconds.\footnote{The 
QSO was observed for 10743 seconds on December 4 ({\it HST} archive ID 
no. O4X002010) and for 13735 seconds on December 11 ({\it HST} archive 
ID nos. O4X001010 and O4X001020).} The observations were obtained with the 
medium resolution FUV echelle mode (E140M) with the 0.2$\times$0.2'' 
aperture. Kimble et al.\markcite{kim98} (1998) report that this STIS 
mode provides a resolution of $R \ = \ \lambda /\Delta \lambda$ = 46000 
(FWHM $\approx$ 7 \kms ) with the 0.2$\times$0.06'' slit. With the 
0.2$\times$0.2'' slit, the FWHM is nearly identical but the broad 
wings of the line spread function are stronger (see 
Figure 13.87 in the Cycle 9 STIS Instrument Handbook). Photoevents were 
recorded by the FUV-MAMA detector in accumulation mode with on-board 
orbital Doppler compensation, and individual exposures ranged from 2215 
to 2880 seconds in duration. Short exposures of a wavelength 
calibration lamp were obtained with each individual observation of the 
QSO. 

The data were reduced with the software developed by the STIS 
Instrument Definition Team (IDT). Pixel-to-pixel sensitivity variations 
in the individual exposures were first corrected with a postlaunch 
flatfield, and then the individual images were aligned by 
cross-correlation\footnote{The images can be aligned by 
cross-correlating the target images or the corresponding wavelength 
calibration lamp images. In principle direct cross-correlation of the 
target images is preferred because drift of the target in the aperture 
could cause shifts derived from the comparison lamp exposures to differ 
from the shifts appropriate for the target exposures. However, in this 
case the target images must contain enough sharp features to produce a 
sharp peak in the cross-correlation. We tried both methods, and we 
found that for this QSO observation, slightly better resolution was 
always obtained by cross-correlating the wavelength lamp exposures to 
determine the image shifts.} and then coadded. To prolong the lifetime 
of the MAMA detector, the position of the dispersed spectrum on the 
detector is changed every few weeks. This was done between the first 
observation of PG 0953+415 and the second, which is beneficial because 
when the data are aligned and coadded, any residual fixed-pattern noise 
not adequately removed by the flatfield is reduced. This also enables 
the identification of spurious features by comparison of the 
wavelength-calibrated spectra obtained on the two occasions. After 
coaddition of the individual images, the scattered light correction 
described by Bowers and Lindler\markcite{bl99} (1999) was applied. Then 
the spectra from each order were extracted with a simple unweighted 
slit with the standard height of 11 ``lores'' pixels,\footnote{The 
MAMAs are designed to support half-pixel centroiding (see 
Timothy\markcite{tim94} 1994), and these half-pixels are usually 
referred to as ``hires'' pixels. This capability was not employed here 
since it does not enhance the data for this particular observation.} 
and smoothing of the background region also followed standard 
procedures for the STIS echelle modes. During the extraction, obviously 
hot pixels were fixed by interpolation between the adjacent pixels 
in the dispersion direction. The wavelength scale was 
computed with a postlaunch dispersion relation (which is updated when
the position of the spectrum on the detector is changed) and put on a 
heliocentric basis. Finally, the data were flux calibrated 
and overlapping regions of adjacent orders were coadded with weighting 
based on signal-to-noise (S/N). The wavelength 
coverage is $\sim$1150-1730 \AA\ with 5 small gaps between orders at 
$\lambda \ >$ 1630 \AA .

\begin{figure}
\plotfiddle{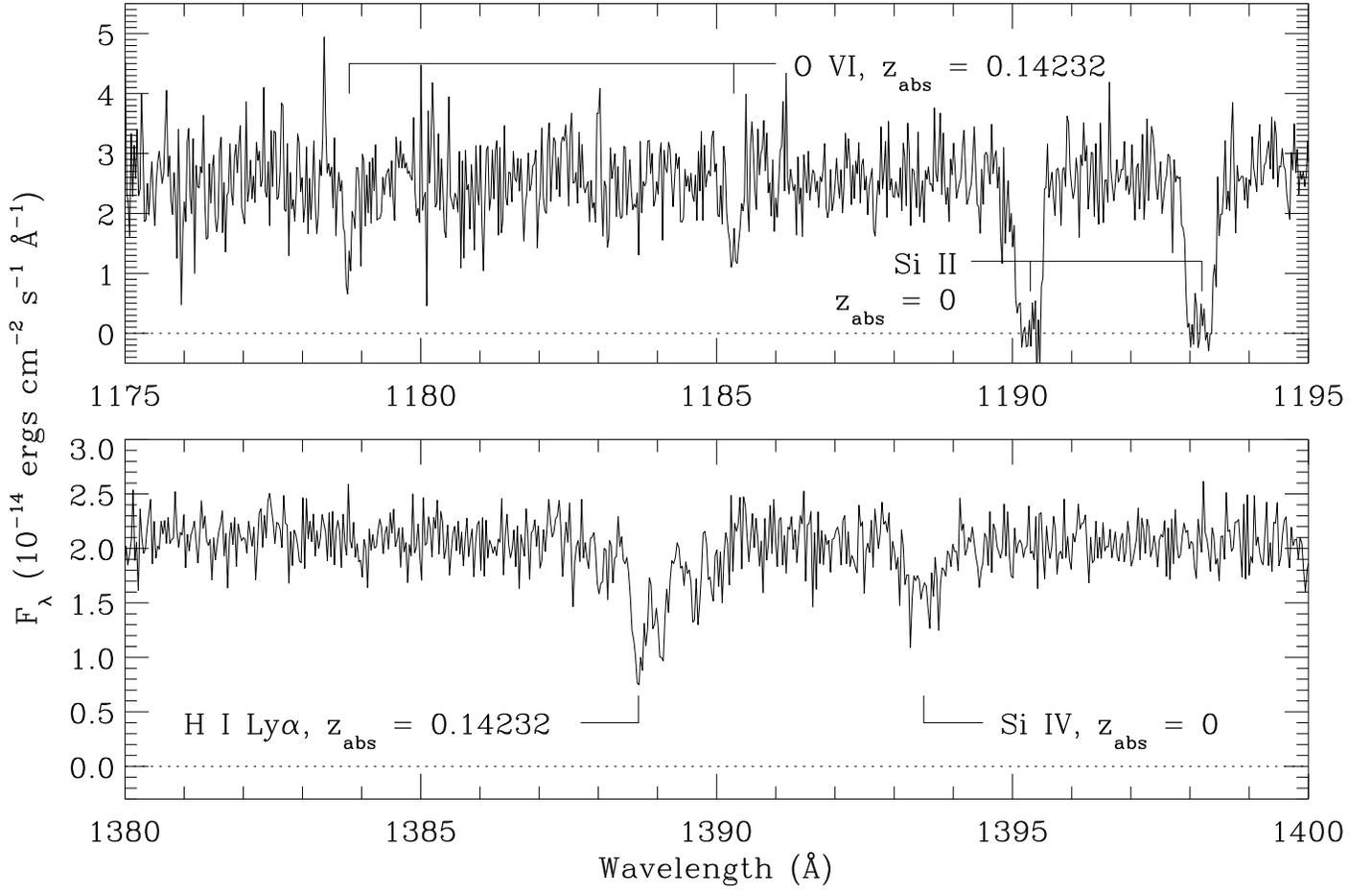}{4in}{90}{85}{85}{350}{0}
\caption[]{Portions of the STIS spectrum of PG 0953+415 obtained with the
intermediate resolution FUV echelle mode (E140M). The \ion{O}{6}
$\lambda \lambda$1032,1038 lines and the \ion{H}{1} Ly$\alpha$ line at
$z_{\rm abs}$ = 0.14232 are marked. Unrelated Milky Way lines in the
selected regions are also indicated. In this figure, the data have been
binned two pixels into one for display purposes only (i.e., all measurements
in the tables and text were made using the full resolution unbinned
data).\label{specsample}}
\end{figure}

Samples of the final spectrum are shown in Figure ~\ref{specsample}.
We identify the lines at 1178.8 and 1185.3 \AA\ as the \ion{O}{6} 
$\lambda \lambda$1032,1038 doublet at \zabs\ = 0.14232, and the 
corresponding \ion{H}{1} \lya line is well-detected at 1388.7 \AA .
A marginal absorption line is apparent at the expected wavelength of
Ly$\beta$ as well, but it is recorded in a region where the spectrum is quite 
noisy.  Note that the \lya profile shows several components, the strongest of
which is well-aligned with the \ion{O}{6} lines. There are no convincing
alternative identifications of these lines. They are unlikely to be 
due to the ISM of the Milky Way since there are no resonance absorption 
lines at these wavelengths (see Morton\markcite{mort91} 1991). The lines at
1178.8 and 1185.3 \AA\ cannot be Ly$\beta , \gamma ,$ or $\delta$ lines due 
to extragalactic absorption systems at different redshifts because the 
corresponding \lya lines are not detected at the expected wavelengths. 
Similarly, we have searched for alternative metal line identifications at 
other redshifts, and we find no convincing candidates.

\begin{figure}
\plotone{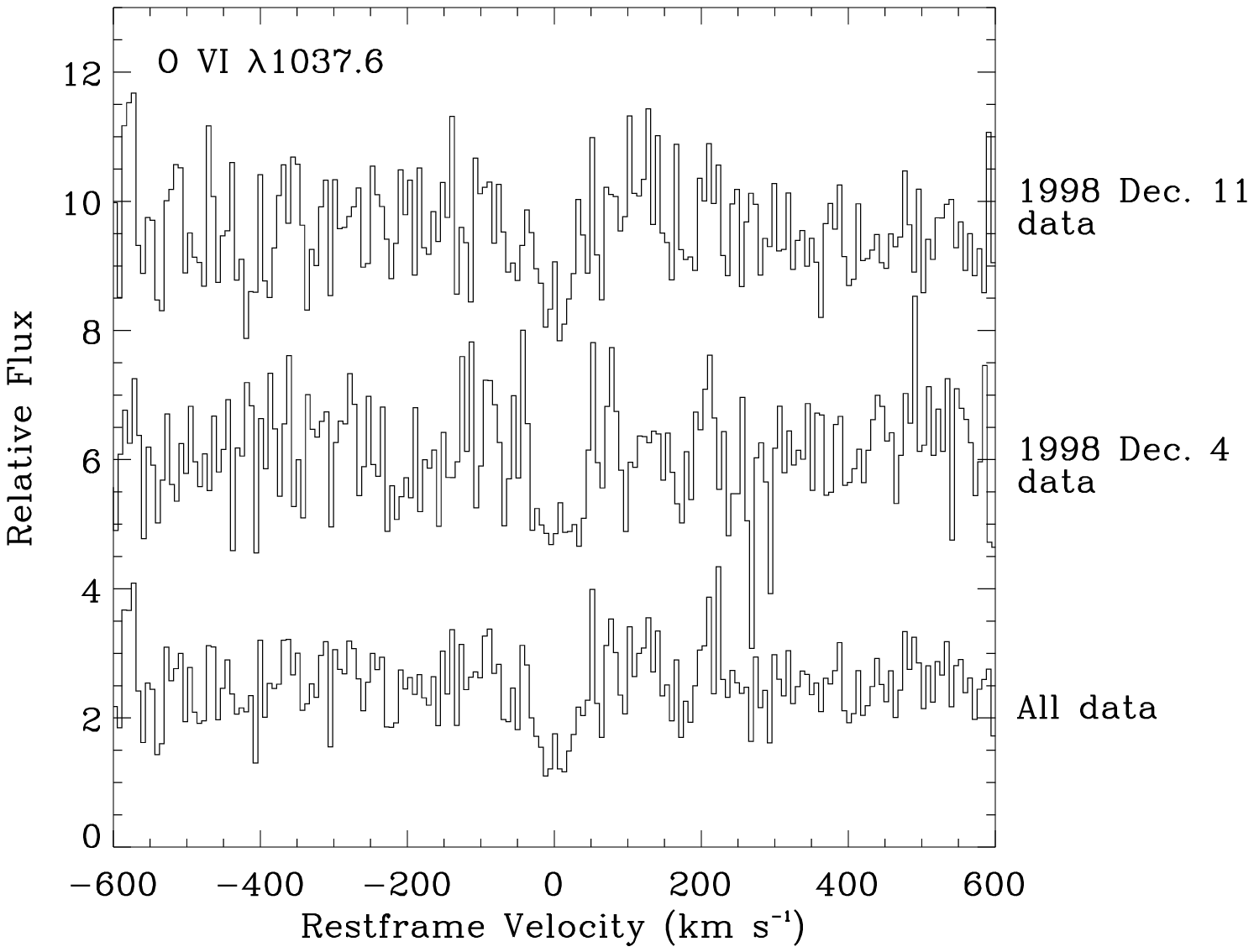}
\caption{Comparison of the \ion{O}{6} $\lambda$1037.6 profile at
\zabs\ = 0.14232 in the data obtained on 1998 December 11 (upper
spectrum), on 1998 December 4 (middle spectrum), and in all of the
data combined (lower spectrum). The spectra are plotted versus
restframe velocity where $v$ = 0 at \zabs\ = 0.14232 and are offset for
clarity. As in Figure 1, the data have been binned two pixels into
one.\label{compare}}
\end{figure}

Since the \ion{O}{6} doublet was recorded in a spectral region which is 
relatively noisy, and since there are slightly hot spurious pixels 
apparent in this portion of the spectrum (see Figure 
~\ref{specsample}), we inspected the spectra obtained on December 4 and 
December 11 independently to confirm that the \ion{O}{6} lines are 
real. Both \ion{O}{6} lines are detected in the individual December 4 
and December 11 spectra, and the relative line depths and wavelength 
difference of the lines are consistent with the \ion{O}{6} 
identification. However, we noticed that the \ion{O}{6} $\lambda 
1037.6$ line is broader in the December 4 spectrum than 
the December 11 spectrum. For the reader's inspection, Figure ~\ref{compare} 
compares the \ion{O}{6} $\lambda$1037.6 profiles derived from the 1998 
December 4 and 1998 December 11 data and from all of the data combined. 
We suspect that the 1037.6 \AA\ line is broader in the December 4 spectrum due 
to noise, perhaps due to fixed pattern noise which was not adequately 
removed by the flatfield available when the data were reduced. Of course, 
this impacts the final spectrum as well, and the 1037.6 \AA\ line is 
broader than the 1031.9 \AA\ line in the final spectrum shown in 
Figure ~\ref{specsample}, though the magnitude of the discrepancy is 
reduced. It is alternatively possible that the \ion{O}{6} $\lambda$1031.9 
line is artificially narrow due to noise or an unrecognized warm pixel 
which fills in the profile somewhat -- we cannot rule out or favor any of 
these possibilities with the current STIS data. Consequently, we provide below 
results from measurement of the 1031.9 \AA\ line only as well as results 
from measurement of both \ion{O}{6} lines.

\subsection{WIYN Galaxy Redshift Measurements}

As part of our continuing study of the relationship between low $z$ 
\lya absorbers and galaxies (see Tripp, Lu, \& Savage\markcite{tls98} 
1998), we have also measured the redshifts of galaxies in the $\sim 
1^{\circ}$ field centered on PG 0953+415 with the fiber-fed multiobject 
spectrograph (Hydra, see Barden \& Armandroff\markcite{ba95} 1995) on 
the WIYN telescope. The observations and measurement techniques are 
described in Tripp et al.\markcite{tls98} (1998), and the new redshifts 
will be provided in a subsequent paper. Here we note that this survey 
has revealed four galaxies within $\sim$130 \kms\ of the \ion{O}{6} 
absorber at \zabs\ = 0.14232. The measured redshifts of these galaxies 
are summarized in Table ~\ref{galtab} along with some of their 
properties including projected distance from the sight line ($\rho$), 
velocity displacement from the absorber, and $O$ and $E$ magnitudes 
from the revised APS digitization of the POSS I plates (Pennington et 
al.\markcite{pen93} 1993). We also list $B_{J}$ magnitudes calculated 
using the relation between $O_{\rm APS}$ and $B_{J}$ derived by Odewahn 
\& Aldering\markcite{oa95} (1995) and the corresponding absolute magnitudes 
in Table ~\ref{galtab}. These are all luminous galaxies ranging from 
0.6$L*$ to 4.0$L*$ (assuming $M_{B}*$ = --19.5 from Loveday et 
al.\markcite{love92} 1992). The velocity dispersion of this group is
very uncertain since we do not know the redshift of the group center of
mass and because of the small sample.
However, if we assume that the \ion{O}{6} absorption arises at the center
of mass of the group, then the radial velocity dispersion is roughly 100 \kms ,
consistent with typical values observed in poor groups (e.g., Zabludoff \&
Mulchaey\markcite{zab} 1998).

\begin{deluxetable}{ccccccccc}
\tablewidth{0pc}
\tablecaption{Galaxies within $\sim$130 km s$^{-1}$ of the \ion{O}{6}
Absorber\label{galtab}}
\tablehead{Galaxy & $\rho$\tablenotemark{b} & $\Delta
v$\tablenotemark{c} & R.A.\tablenotemark{d} & Decl.\tablenotemark{d} &
$O_{\rm APS}$\tablenotemark{e} & $(O - E)_{\rm APS}$\tablenotemark{e} &
$B_{J}$\tablenotemark{f} & $M_{B}$\tablenotemark{g} \nl
Redshift\tablenotemark{a} & (Mpc) & (km s$^{-1}$) & (J2000) & (J2000) & \ & \ &
\ &}
\startdata
0.14280 & 0.395 & 126 & 9 56 38.90 & 41 16 46.6 & 18.4 & 1.3 & 17.9 &
--21.0 \nl
0.14282 & 1.00  & 131 & 9 57 21.41 & 41 20 17.3 & 19.0 & 1.4 & 18.5 &
--20.4 \nl
0.14257 & 2.38  & 66 & 9 58 15.07 & 41 23 23.8 & 20.4 & 2.5 & 20.0 &
--18.9 \nl
0.14274 & 3.00  & 110 & 9 55 14.17 & 41 3 23.3 & 19.1 & 0.6 & 18.6 &
--20.3
\enddata
\tablenotetext{a}{Internal redshift errors are estimated to be $\sim$50 \kms .}
\tablenotetext{b}{Projected distance to the sight line (impact
parameter) assuming $H_{0}$ = 75 km s$^{-1}$ Mpc$^{-1}$ and $q_{0}$ =
0.0. The QSO coordinates (J2000) are R.A. = 9$^{\rm h} 56^{\rm m}52\fs
4$, decl. = +41$^{\rm d} 15\arcmin 22\farcs 0$.}
\tablenotetext{c}{$\Delta v \ = \ c(z_{\rm gal} - z_{\rm abs})/(1 +
z_{\rm mean})$ where $z_{\rm mean}$ is the mean of $z_{\rm abs}$ and
$z_{\rm gal}$.}
\tablenotetext{d}{Units of right ascension are hours, minutes, and
seconds, and units of declination are degrees, arcminutes, and
arcseconds.}
\tablenotetext{e}{POSS I $O$ and $E$ magnitudes from the revised APS
catalog (Pennington et al.\markcite{pen93} 1993), see
http://aps.umn.edu.}
\tablenotetext{f}{B magnitude calculated from the relation derived by
Odewahn \& Aldering\markcite{oa95} (1995): $O_{\rm APS} -18 =
0.96(B_{J} - 18) + 0.52$.}
\tablenotetext{g}{Absolute magnitude calculated using the interstellar
extinction correction based on $E(B-V)$ from Lockman \& Savage 1995 and
the $K$--correction $K \ = \ 2.5 \ {\rm log} \ (1+z)$.}
\end{deluxetable}

The spectrum of the galaxy in Table ~\ref{galtab} with the smallest 
projected separation from the sight line ($\rho$ = 395 kpc) shows 
strong Balmer lines in absorption as well as the other usual absorption 
lines, e.g., \ion{Ca}{2} H \& K, and well detected [\ion{O}{2}] and 
H$\beta$ emission lines with observed equivalent widths of 10.5$\pm$2 
\AA\ and 4.5$\pm$0.5 \AA , respectively. However, the [\ion{O}{3}] 
emission lines are not apparent (H$\alpha$ is redshifted beyond the red 
end of the spectrum). Based on its luminosity and the 
Kennicutt\markcite{ken92a} \markcite{ken92b} (1992a,b) atlas of 
integrated galaxy spectra, this implies that the galaxy is a normal 
Sb-Sc spiral or a peculiar S0-Sa galaxy, without an active nucleus. The 
galaxy at $\rho$ = 3.0 Mpc is also an emission line galaxy with 
[\ion{O}{2}], [\ion{O}{3}], H$\beta$, and even H$\gamma$ seen in 
emission. This object has strong emission lines; the observed 
[\ion{O}{2}] equivalent width is 37$\pm$10 \AA . There are no 
indications of an active nucleus (e.g., broad H$\beta$ or large 
[\ion{O}{3}]/H$\beta$ ratios), so given its high luminosity we classify 
this galaxy as a starbursting spiral or a normal Sc-Sd spiral. The 
other galaxies in Table ~\ref{galtab} do not have emission lines and 
cannot be unambiguously classified.

It is important to note that the galaxy redshift survey we have been able to 
carry out to date is not very complete except at relatively bright limiting 
magnitudes, and it is quite possible that there are additional galaxies 
at $z \approx$ 0.142 which are closer to the sight line than those in 
Table ~\ref{galtab}. Consequently, we cannot comment on whether or not
a dwarf galaxy could give rise to the \ion{O}{6} system, for example. 
As we shall discuss in \S 5.1, it is difficult to discriminate between the 
hypothesis that the \ion{O}{6} absorption originates in an intragroup
medium and the hypothesis that it occurs in the ISM of an individual 
galaxy given the limited information currently available.  More galaxy 
redshift measurements and deep imaging would be valuable.

\section{Absorption Line Measurements}

\begin{figure}
\plotfiddle{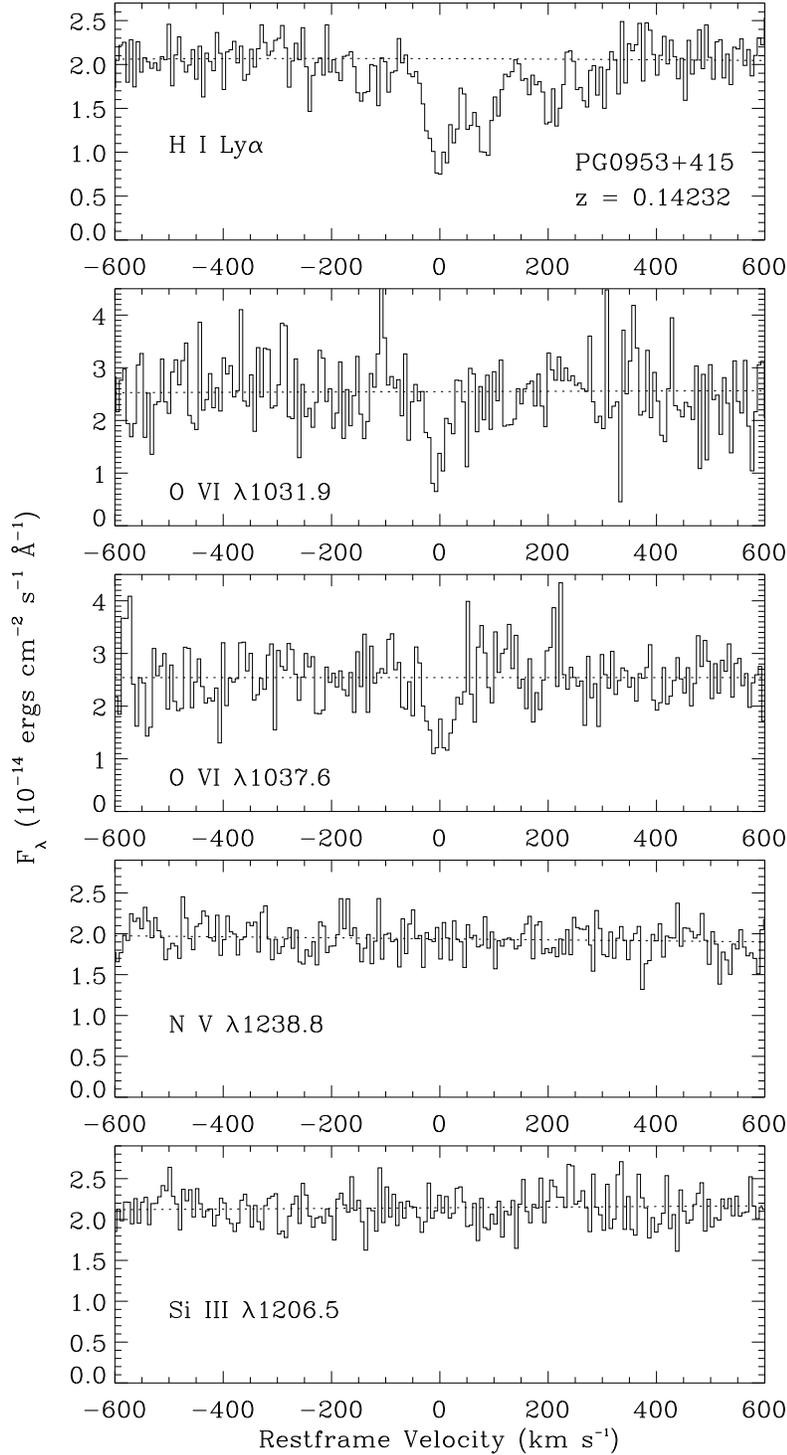}{7.0in}{0}{72}{72}{-200}{-50}
\caption[]{Absorption line profiles of the \ion{O}{6} and \ion{H}{1} \lya
lines at $z_{\rm abs}$ = 0.14232, plotted versus restframe velocity.
The dotted lines show the continua adopted for equivalent width and apparent
column density measurements as well as profile fitting (see text \S 3).
Spectral regions of lines of interest which are {\it not}
detected are shown in the lower panels. In this figure, the data have been
binned two pixels into one (for display purposes only).\label{profiles}}
\end{figure}

The absorption line profiles of the \ion{O}{6} and \ion{H}{1} \lya 
lines at $z_{\rm abs}$ = 0.14232 are plotted versus restframe velocity 
in Figure ~\ref{profiles}. This figure also shows the spectral regions 
of lines of interest which are not detected. Restframe 
equivalent widths ($W_{\rm r}$) of the \ion{O}{6} and \ion{H}{1} 
absorption lines were measured using the software of Sembach \& 
Savage\markcite{ss92} (1992), but adapted to use the statistical 
uncertainties directly calculated from the signal and background counts 
recorded by STIS. This software accounts for errors due to uncertainty 
in the height and curvature of the continuum as well as a 2\% uncertainty 
in the flux zero point in the overall uncertainty in 
$W_{\rm r}$ [see the appendix in Sembach \& Savage\markcite{ss92} 
(1992) for details]. These equivalent widths are listed in Table 
~\ref{lineprop}. This software was also used to set upper limits on the 
\ion{N}{5}, \ion{Si}{3}, \ion{Si}{2}, and \ion{C}{2} equivalent widths, 
which are also listed in Table ~\ref{lineprop}.

\begin{deluxetable}{lccccc}
\footnotesize
\tablewidth{0pc}
\tablecaption{Equivalent Widths and Integrated Column Densities of the
\ion{O}{6} Absorber at z = 0.14232\label{lineprop}}
\tablehead{Species & $\lambda _{0}$\tablenotemark{a} & $W_{\rm
r}$\tablenotemark{b} & log $N_{\rm a}$\tablenotemark{c} & $v_{-}$ &
$v_{+}$ \nl
 \ & (\AA ) & (m\AA ) & \ & (km s$^{-1}$) & (km s$^{-1}$) }
\startdata
\ion{H}{1} & 1215.67 & 423$\pm$28 & 14.00$\pm$0.03 & --175 & 330 \nl
\ion{O}{6} & 1031.93 & 84$\pm$15  & 14.04$^{+0.10}_{-0.13}$ & --40 & 40
\nl
\ion{O}{6} & 1037.62 & 97$\pm$15  & 14.35$^{+0.08}_{-0.10}$ & --40 & 40
\nl
\ion{N}{5} & 1238.80 & $<$36\tablenotemark{d} &
$<$13.2\tablenotemark{d} & --40 & 40 \nl
\ion{Si}{4} & 1393.76 & $<$69\tablenotemark{d} &
$<$12.9\tablenotemark{d} & --40 & 40 \nl
\ion{Si}{3} & 1206.50 & $<$41\tablenotemark{d} &
$<$12.3\tablenotemark{d} & --40 & 40 \nl
\ion{Si}{2} & 1304.37\tablenotemark{e} & $<$27\tablenotemark{d} &
$<$13.3\tablenotemark{d} & --40 & 40 \nl
\ion{C}{2} & 1334.53 & $<$42\tablenotemark{d} &
$<$13.3\tablenotemark{d} & --40 & 40 \nl
\ion{C}{4} & 1548.20 & $<$280\tablenotemark{f} &
$<$13.8\tablenotemark{g} & \nodata & \nodata
\enddata
\tablenotetext{a}{Restframe vacuum wavelength from
Morton\markcite{mort91} (1991). Oscillator strengths used for these
measurements were also obtained from Morton\markcite{mort91} (1991).}
\tablenotetext{b}{Restframe equivalent width integrated from $v_{-}$ to
$v_{+}$. This velocity range includes all readily apparent components
(see Figure 2).}
\tablenotetext{c}{Apparent column density (see text) integrated from
$v_{-}$ to $v_{+}$.}
\tablenotetext{d}{4$\sigma$ upper limit.}
\tablenotetext{e}{The strongest Si II line at 1260.42 \AA\ is
recorded in a region of the spectrum affected by the FUV MAMA repeller
wire, so we elected to use the Si II $\lambda$1304.37 line instead. In
the wavelength range covered by the STIS spectrum, the Si II lines
at 1190.42 and 1193.29 \AA\ are expected to be stronger than the 1304.37
\AA\ line, but these lines have unreliable oscillator strengths (see
\S 3.2.1 in Tripp, Lu, \& Savage\markcite{tls96} 1996).}
\tablenotetext{f}{4$\sigma$ upper limit from the Faint Object
Spectrograph observations reported by Jannuzi et al.\markcite{jan98} (1998).}
\tablenotetext{g}{The 4$\sigma$ upper limit for log $N$(C IV) obtained
assuming unsaturated absorption.}
\end{deluxetable}

\begin{deluxetable}{lcccc}
\small
\tablewidth{0pc}
\tablecaption{Velocities, Doppler Parameters, and Column Densities of
Individual Components of the z = 0.14232 Absorber\label{compprop}}
\tablehead{Species & $\lambda _{0}$\tablenotemark{b} &
$v$\tablenotemark{c} & $b$ & log $N$ \nl
\ & (\AA ) & (km s$^{-1}$) & (km s$^{-1}$) & \ }
\startdata
\ion{H}{1} & 1215.67 & --142$\pm$3 & 12$^{+15}_{-7}$     & 12.74$\pm$0.10 \nl
   \       & \       & 1$\pm$2     & 31$\pm$7            & 13.59$\pm$0.03 \nl
   \       &   \     & 81$\pm$2    & 30$\pm$9            & 13.43$\pm$0.04 \nl
   \      &   \    & 163$\pm$7   & $\sim$20\tablenotemark{d} & 12.6$\pm$0.21\nl
   \       &   \     & 207$\pm$3   & 19$^{+15}_{-9}$    & 13.08$\pm$0.08 \nl
   \       &   \     & 269$\pm$4   & 17$^{+26}_{-10}$    & 12.81$\pm$0.12 \nl
\ion{O}{6} & 1031.92\tablenotemark{e} & -3$\pm$3    & 19$^{+13}_{-8}$     &
14.03$\pm$0.09 \nl \hline
\ion{O}{6} & doublet\tablenotemark{f} & -2$\pm$2    & 22$^{+10}_{-7}$     &
14.16$\pm$0.06
\enddata
\tablenotetext{a}{Individual component parameters determined from profile
fitting with the software of Fitzpatrick \& Spitzer\markcite{fitz97} (1997).}
\tablenotetext{b}{Restframe vacuum wavelength from
Morton\markcite{mort91} (1991). Oscillator strengths used for these
measurements were also obtained from Morton\markcite{mort91} (1991).}
\tablenotetext{c}{Velocities are in the restframe of the \ion{O}{6}
absorber with $v$ = 0 km s$^{-1}$ at \zabs\ = 0.14232.}
\tablenotetext{d}{This parameter is poorly constrained by the profile fitting
software; the nominal value at the minimum $\chi ^{2}$ is listed.}
\tablenotetext{e}{Component parameters based on fitting of the \ion{O}{6}
$\lambda$1031.92 line {\it only}. For the reasons noted in \S 2.1, this fit
is preferred over the fit to both lines of the \ion{O}{6} doublet.}
\tablenotetext{f}{Component parameters based on fitting of {\it both} lines of
the \ion{O}{6} $\lambda \lambda$1031.92, 1037.62 doublet.}
\end{deluxetable}

To measure column densities, we have used two methods: the apparent 
column density technique (e.g., Savage \& Sembach\markcite{ss91} 1991) 
and Voigt profile fitting. For Voigt profile fitting, we have used the 
software of Fitzpatrick \& Spitzer\markcite{fitz97} (1997) with the line
spread functions for the E140M mode with the 0.2$\times$0.2'' aperture 
shown in Figure 13.87 of the Cycle 9 STIS Handbook. This software provides 
Doppler parameters ($b$), velocities ($v$), and column densities ($N$) for a 
number of components which is subjectively specified by the user. The 
profile fitting results are summarized in Table ~\ref{compprop}.
Since the \ion{O}{6} $\lambda$1037.6 \AA\ 
line may be artificially broadened by a noise feature (see \S 2.1), we present
two alternative fits in Table ~\ref{compprop}: a fit to the \ion{O}{6} 
$\lambda$1031.9 line {\it only} and a fit to both of the
\ion{O}{6} lines. 

In the apparent column density approach, the apparent column density 
per unit velocity, $N_{\rm a} (v)$, is calculated directly from the 
apparent optical depth,
\begin{equation}
N_{\rm a}(v) \ = \ (m_{\rm e}c/\pi e^{2})(f\lambda )^{-1}\tau _{\rm a}(v) 
\ = \ 3.768 \times 10^{14} (f\lambda )^{-1} {\rm ln}[I_{\rm c}(v)/I(v)]
\end{equation}
in atoms cm$^{-2}$ (\kms )$^{-1}$, where $f$ is the oscillator 
strength, $\lambda$ is the wavelength in \AA , $I(v)$ is the observed 
line intensity and $I_{\rm c} (v)$ is the estimated continuum intensity 
at velocity $v$, and the other symbols have their usual meanings. If 
the lines are fully resolved or are not affected by saturation, then 
the total column density can be obtained by simple integration, $N = 
\int N_{\rm a}(v) dv$. We used the software of Sembach \& 
Savage\markcite{ss92} (1992) to measure the apparent column densities, 
so again, uncertainties due to the continuum placement and a 2\% 
uncertainty in the flux zero point are included in the final error bars.  

\begin{figure}
\plotfiddle{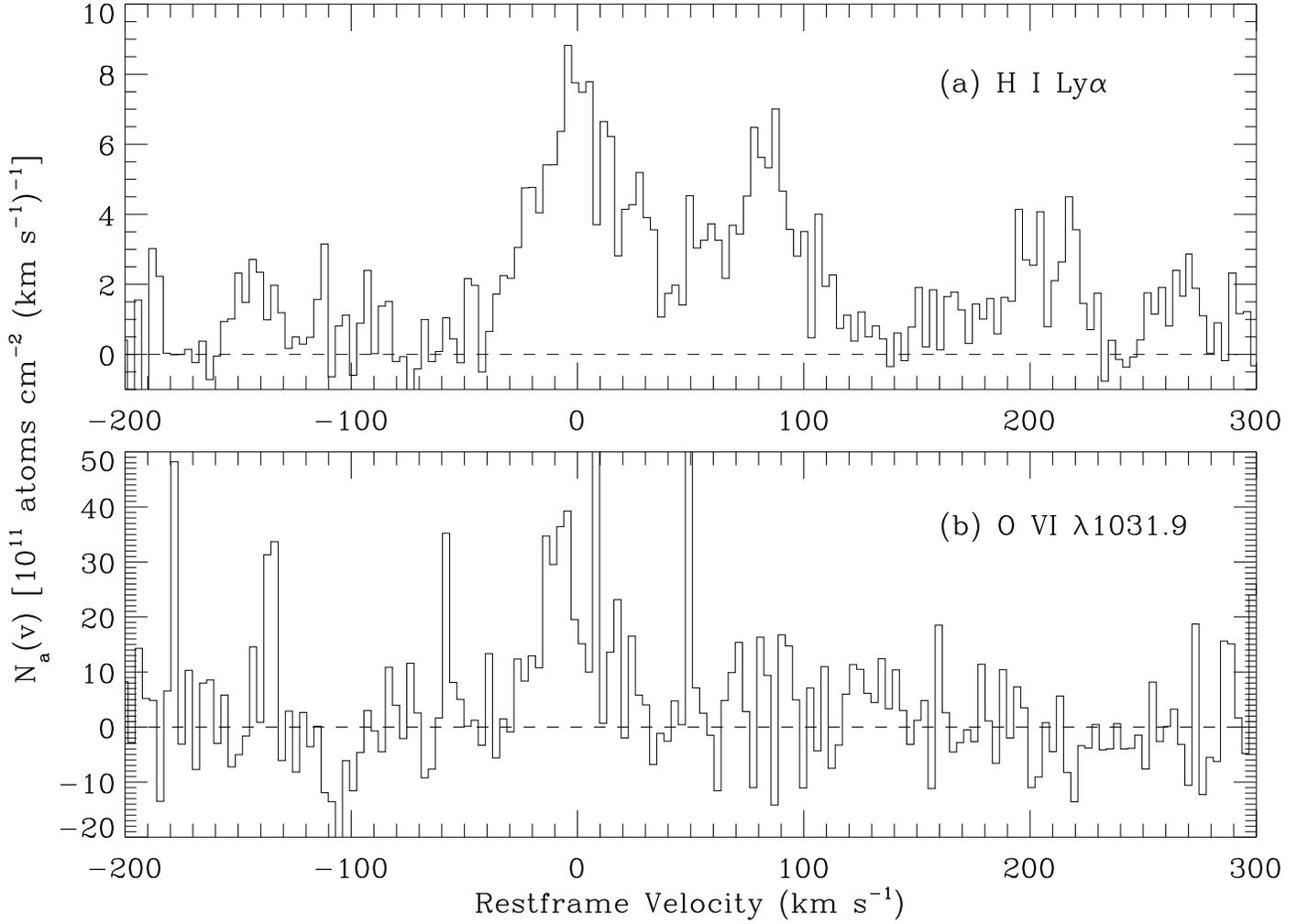}{4in}{90}{85}{85}{350}{0}
\caption[]{Apparent column density profiles (see text \S 3) of (a) the
\ion{H}{1} \lya and (b) \ion{O}{6} $\lambda$1031.9 lines plotted versus
restframe velocity where $v$ = 0 at \zabs\ = 0.14232.
There appears to be a small difference between the
velocity centroids of the \ion{O}{6} and the main component of the \ion{H}{1}
profile. However, the component velocities from profile fitting agree within
the 1$\sigma$ uncertainties (see Table ~\ref{compprop}). These \nav\ profiles
are constructed from the full resolution, i.e., unbinned, data.\label{nav}}
\end{figure}

The \ion{H}{1} \lya and \ion{O}{6} $\lambda$1031.9 \nav\ profiles are 
plotted in Figure ~\ref{nav}, and their integrated apparent column densities
are provided in Table ~\ref{lineprop}.  This table also lists $4\sigma$ upper 
limits on column densities of several species of interest obtained by 
integrating \nav\ over the velocity range of the undetected lines. 
We also include a $4\sigma$ upper limit on \ion{C}{4} $\lambda$1548.2 from the
Faint Object Spectrograph (FOS) observations of Jannuzi et 
al.\markcite{jan98} (1998). Figure ~\ref{nav} 
shows that there appears to be a small velocity difference between the 
centroid of the strongest component of the \ion{H}{1} profile and that of
the \ion{O}{6} line. However, the \ion{H}{1} and \ion{O}{6} velocities from 
profile fitting agree within their 1$\sigma$ uncertainties.  More importantly, 
Figure ~\ref{nav} shows that the main component of the \ion{H}{1} profile 
is broader than the \ion{O}{6} 1031.9 \AA\ line, which suggests that the line 
widths are (at least partially) controlled by thermal motions (see \S 4.1). 
However, it is alternatively possible that the two strongest components of the 
\ion{H}{1} profile include contributions from narrow features which are not 
detected in the \ion{O}{6} line. Such additional narrow lines, if present, 
are not well-constrained by the current data.

\section{Physical Conditions}

We now turn to the physical conditions of the absorbing gas. As discussed in
\S 1, we are especially interested in testing the warm/hot gas prediction of
cosmological simulations of structure growth. For this purpose we seek to
constrain the gas temperature in \S 4.1, and we examine the ionization 
mechanism in \S 4.2.

\subsection{Gas Temperature}

If the line broadening is entirely due to thermal motions, then the gas 
temperature can be determined from the Doppler parameter of a given line,
\begin{equation}
T = mb^{2}/2k = A(b/0.129)^{2},
\end{equation}
where $m$ is the mass and $A$ is the atomic mass number of the element, the 
Doppler parameter $b$ is in \kms , and $T$ is in $^{\circ}$K. However, it 
is usually true that additional factors such as gas turbulence or multiple 
unresolved components contribute to the width of a line, and consequently 
the gas temperature from equation (2) must be treated as an upper limit. 
If lines of two or more elements with adequately different masses are 
available, then $b$ can be expressed as
\begin{equation}
b^{2} = b_{nt}^{2} + (0.129)^{2}T/A,
\end{equation}
which can be solved for $T$ and $b_{nt}$, the component of the broadening 
due to non-thermal motions, assuming the different absorption lines arise
in the same gas and that the non-thermal motions (turbulence) can be 
assumed to have a Gaussian profile.

The temperature upper limits implied by equation (2) and the Doppler 
parameters in Table ~\ref{compprop} are $T$(\ion{O}{6}) $\leq \ 3.5\times 
10^{5} \ ^{\circ}$K and $T$(\ion{H}{1}) $\leq \ 5.8\times 10^{4} \ ^{\circ}$K
for the strongest component of the \ion{H}{1} profile, which is also 
the closest component to the \ion{O}{6} line in velocity space (see Figure 
~\ref{nav}). Since the velocity centroids of the \ion{O}{6} and the 
\ion{H}{1} are quite close and, in fact, in agreement within the 1$\sigma$
error bars, it is plausible that these \ion{O}{6} and \ion{H}{1} lines 
originate in the same gas. Adopting this assumption, we derive from 
equation (3) $T$ = $3.9\times 10^{4} \ ^{\circ}$K and $b_{nt}$ = 18 \kms .

While such low temperatures clearly favor photoionization (see below), we
must bear in mind that there are considerable uncertainties in the Doppler
parameters.  For example, for the strongest component of the \ion{H}{1} 
profile, increasing the $b$-value in Table ~\ref{compprop} by just 1$\sigma$
increases the upper limit on $T$(\ion{H}{1}) to $8.7 \times 10^{4} \ 
^{\circ}$K, much closer to the temperature range where \ion{O}{6} can 
be produced by collisional ionization. This issue is exacerbated by the
non-uniqueness of the fitted profile when dealing with multiple blended
components, i.e., assuming a different number of components or a different
mix of broad and narrow components can lead to substantially different
$b$-values. We provide an example of this in \S 4.2.2.  It is also possible
that the absorption arises in non-equilibrium collisionally ionized gas
in which cooling is faster than recombination (see \S 4.2.2).

\subsection{Ionization}

Given the measurements in the previous sections, can we favor 
photoionization or collisional ionization in this \ion{O}{6} absorber?
In the Milky Way ISM, \ion{O}{6} likely traces collisionally ionized
hot gas (Jenkins\markcite{jenk78a}\markcite{jenk78b} 1978a,b), at least 
in the disk and lower halo.  However, 
in extragalactic \ion{O}{6} absorbers, the EUV ionizing radiation field may be 
substantially harder, and the absorption might arise in very low density gas 
with a long path length. Both of these factors make photoionization more viable 
in the extragalactic case.  We first consider the plausibility of photoionized 
models (\S 4.2.1), and then we test the collisional scenario (\S 4.2.2).

\subsubsection{Photoionization}

We have explored the photoionized scenario using CLOUDY (version 90.04; Ferland 
et al.\markcite{fer98} 1998) with the standard assumptions, in particular that
there has been time for thermal and ionization equilibrium to prevail (see 
below). The absorber is treated as a constant density plane-parallel slab 
photoionized by the extragalactic radiation from QSOs and AGNs, as calculated 
by Haardt \& Madau\markcite{hm96} (1996) for $z \approx$ 0.12. 
We set the mean intensity at the \ion{H}{1} Lyman limit to $J_{\nu }$(LL) = 
1 $\times \ 10^{-23}$ ergs s$^{-1}$ cm$^{-2}$ Hz$^{-1}$ sr$^{-1}$, a value 
in agreement with observational constraints (e.g., Kulkarni \& 
Fall\markcite{kf93} 1993; Maloney\markcite{mal93} 1993; Tumlinson et al.
\markcite{tum99} 1999) and theoretical expectations (Shull et al.
\markcite{shu99} 1999). With these assumptions, we varied the 
metallicity\footnote{We express the linear abundance of element X relative to 
element Y as (X/Y) and the logarithmic abundance in the usual fashion, 
[X/Y] = log (X/Y) -- log (X/Y)$_{\odot}$, and we indicate the overall 
metallicity with the variable $Z$.} and ionization parameter 
($U = n_{\gamma }/n_{\rm H}$ = H ionizing photon density/total hydrogen 
number density, neutral + ionized) to search for models which are consistent 
with the constraints set above: the \ion{H}{1} and \ion{O}{6} column 
densities, limits on various column density ratios which can be derived using 
the upper limits in Table ~\ref{lineprop}, and the gas temperature. The 
{\it relative} heavy element abundances (e.g., [N/O]) were initially set to 
the solar values from Grevesse \& Anders\markcite{ga89} (1989) and Grevesse 
\& Noels\markcite{gn93} (1993), then we considered an alternative N/O 
abundance (see below).

\begin{figure}
\plotone{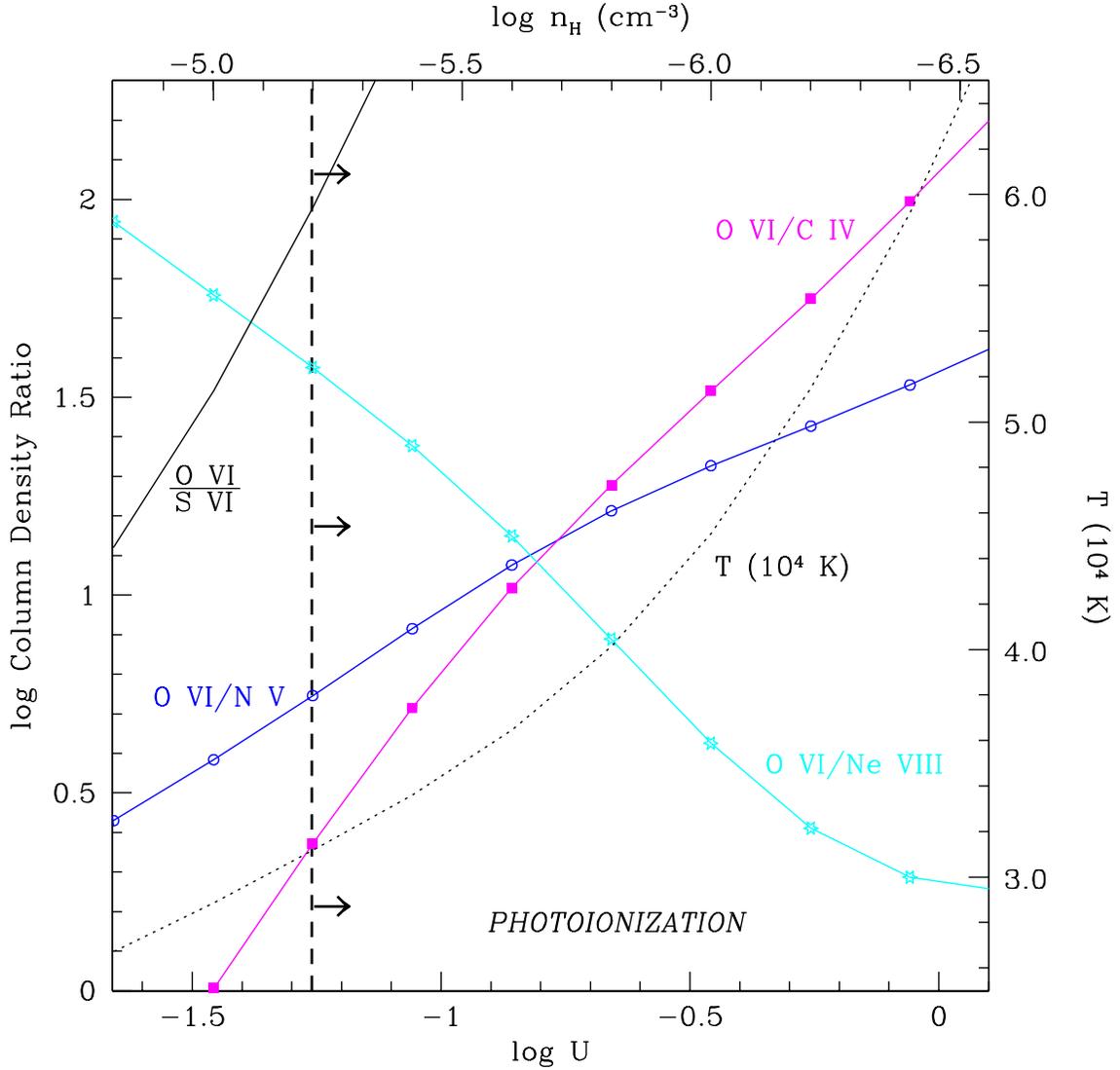}
\caption[]{Photoionization model calculations of logarithmic ratios of the
column densities of \ion{O}{6} to \ion{N}{5} (open circles), \ion{O}{6}
to \ion{C}{4} (filled squares), and \ion{O}{6} to \ion{Ne}{8} (open stars)
plotted versus the ionization parameter (bottom axis) and the hydrogen
number density (top axis). The temperature of the model as a function of
the ionization parameter is also shown with a dotted line using the
linear scale on the right axis (which is in units of $10^{4} \ ^{\circ}$K).
The observational lower limit on $N$(\ion{O}{6})/$N$(\ion{N}{5}) requires
log $U \ \geq$ --1.26, i.e. only the region of the plot to the right of
the vertical dashed line is in agreement with this observational constraint.
\label{photo}}
\end{figure}

The basic results of the photoionization modeling are encapsulated in Figure
~\ref{photo}. This figure shows various column density ratios involving 
\ion{O}{6} and other high ions of interest, plotted as function of the
ionization parameter (bottom axis) and $n_{\rm H}$ (top axis). A metallicity 
of $Z$ = 1/10 $Z_{\odot}$ was used to construct this model, but at low 
metallicities the column densities scale nearly directly with $Z$.\footnote{As
$Z$ approaches the solar value, the increased cooling provided by the 
metals affects $T$ and, in turn, the ionization balance of the gas.} The gas 
temperature of the model as a function of $U$ is also indicated by a dotted 
line in Figure ~\ref{photo} with the linear temperature scale on the 
right axis.  Of the various species that we are able to place limits on using 
the STIS E140M spectrum (see Table ~\ref{lineprop}), only \ion{N}{5} usefully 
constrains the photoionization model assuming [N/O] = 0; when the lower limit 
on $N$(\ion{O}{6})/$N$(\ion{N}{5}) is satisfied, the other species covered
in the STIS E140M spectrum are predicted to have undetectably small column 
densities. Note that the STIS spectrum does not cover the \ion{C}{4} doublet at 
this redshift, and we discuss the constraint set by the FOS upper limit on 
\ion{C}{4} below.  The \ion{O}{6}/\ion{N}{5} column density ratio requires 
log $U \ >$ --1.26. Given this limit on $U$, the photoionized model 
requires $n_{\rm H} \ \leq \ 10^{-5.2}$ cm$^{-3}$ and 
$T \geq$ 31,200 $^{\circ}$K. At this ionization parameter, the \ion{O}{6}
ionization fraction $f$(\ion{O}{6}) = \ion{O}{6}/O$_{\rm total}$ = 0.164,
and therefore the path length $L$ through the constant density absorbing 
region must be greater than $\sim$460 kpc to produce the observed \ion{O}{6} 
column density, log $N$(\ion{O}{6}) = 14.04, in 1/10 solar metallicity gas. 
With such a long path length, expansion of the universe should lead to an
\ion{O}{6} line which is broader than the 1031.9 \AA\ line shown in Figure 
~\ref{nav}, and the photoionized model predicts a substantially 
larger \ion{H}{1} column density than observed. While the narrowness of 
the \ion{O}{6} line may be due to noise, the incorrect $N$(\ion{H}{1}) 
rules out the 1/10 solar metallicity photoionized model. To match the observed 
$N$(\ion{O}{6}) and $N$(\ion{H}{1}) with log $U$ = --1.26 and $J_{\nu}$(LL) 
= $10^{-23}$ ergs$^{-1}$ s$^{-1}$ cm$^{-2}$ Hz$^{-1}$ sr$^{-1}$, the 
photoionization model requires $Z \sim 1.6Z_{\odot}$ and $L \sim$ 25 
kpc, which also alleviates the problem with Hubble broadening and the width
of the line. While this is an uncomfortably high metallicity to require, 
recent {\it ASCA} observations of X-ray bright galaxy groups indicate that 
in some groups, the intragroup gas metallicity is $\sim$1/3--1/2 solar 
(Davis, Mulchaey, \& Mushotzky\markcite{dmm99} 1999; Hwang et 
al.\markcite{hwa} 1999), so this may be marginally plausible. However, 
in such intragroup gas collisional ionization should be important. 

Are these reasonable physical properties? Observations have established that 
some QSO absorbers can be quite extended.  For example, the characteristic 
\lya absorber ``size'' inferred from close QSO pairs is several hundred kpc at 
moderate redshifts (e.g., Dinshaw et al.  \markcite{din95} 1995; Fang et al.
\markcite{fang96} 1996), and this is consistent with cosmological simulations
of structure formation (e.g., Rauch, Haehnelt, \& Steinmetz\markcite{rhs97} 
1997; Dav\'{e} et al.\markcite{dave99} 1999). However, as 
various authors have argued on different grounds, these large
absorbers are likely filamentary structures composed of smaller clouds 
rather than a single monolithic gas cloud (e.g., Rauch, Weymann, \& 
Morris\markcite{rwm96} 1996; Cen \& Simcoe\markcite{cs97} 1997).
Cen \& Simcoe\markcite{cs97} report that individual clouds have sizes of
$\sim$50 kpc. Therefore the path length required by the photoionized model, 
$L \sim$ 25 kpc, is reasonable.
The low densities inferred for the \ion{O}{6} absorber at \zabs\ = 
0.14232 are also consistent with cosmological simulations: according to 
equation (2) in Dav\'{e} et al. (1998), the strongest component of the 
\zabs\ = 0.14232 absorber is expected to have $n_{\rm H} \approx 10^{-5.2}$ 
cm$^{-3}$ based on its measured \ion{H}{1} column density. Such low densities
are also possible in the outer halo of a galaxy like the Milky Way.  
Murali\markcite{mura} (2000) has recently argued that the existence 
of the Magellanic Stream requires $n_{\rm H} \ \lesssim \ 10^{-5}$ cm$^{-3}$ 
in the Milky Way halo at a Galactocentric distance of $\sim$50 kpc.  Evidently,
the size and density of this \ion{O}{6} absorber inferred from the photoionized
model are plausible.

The most problematic property of the photoionized model is the required
metallicity, $Z \sim 1.6 Z_{\odot}$. Most environments in which high 
metallicities are expected should involve collisional ionization or else 
should have substantially higher \ion{H}{1} column densities. If we use
expression (7) from Dav\'{e} et al.\markcite{dave99} (1999) to estimate 
the approximate overdensity associated with the \zabs\ = 0.14232 absorber
based on its observed $N$(\ion{H}{1}), we find that 
$\rho _{\rm H}/\bar{\rho} _{\rm H} \sim$ 20. Based on this overdensity, 
the cosmological simulations of Cen \& Ostriker\markcite{co99b} (1999b) 
predict that the metallicity of this absorber should be $Z \sim 
0.1 Z_{\odot}$. Similarly, to produce the absorption in a high metallicity
region of the ISM of a galaxy like the Milky Way, the \ion{H}{1} column 
density would have to be substantially higher than the observed $N$(\ion{H}{1}).

We note that the H$^{+}$ recombination timescale,
$t_{\rm rec}($H$^{+}) = 1/\alpha (T) n_{\rm e}$ where $\alpha (T)$ is the 
recombination coefficient, exceeds the age of the universe by a factor of 
several at such low densities, and one may wonder if modeling the gas with 
a photoionization equilibrium code is valid.  However, the timescale for the
gas to approach equilibrium, $t_{\rm eq} = [t_{\rm ion}^{-1} + 
2t_{\rm rec}^{-1}]^{-1}$ where $t_{\rm ion}$ is the photoionization timescale 
(see Appendix A in Dove \& Shull\markcite{dove} 1994), is vastly shorter than 
the recombination timescale in the conditions considered here, and 
photoionization equilibrium is a good approximation for the low density, highly 
ionized gas (see, e.g., Vedel, Hellsten, \& Sommer-Larsen\markcite{vedel} 1994).

A caveat in the photoionization models above is that we have assumed the solar 
[N/O] abundance.  If instead
we assume [N/O] $\sim$ --1.5 based on the trend of [N/O] vs. metallicity 
observed in giant extragalactic \ion{H}{2} regions (Vila-Costas \& 
Edmunds\markcite{vce93} 1993), then the lower limit on 
$N$(\ion{O}{6})/$N$(\ion{N}{5}) is satisfied at a lower ionization parameter: 
the model requires log $U >$ --1.90.  However, at lower values of $U$ the 
model predicts that \ion{Si}{3} and \ion{Si}{4} are detectable, and in fact
in this case the lower limit on $N$(\ion{O}{6})/$N$(\ion{Si}{3}) from 
Table ~\ref{lineprop} provides a more stringent constraint requiring
log $U >$ --1.69 and $n_{\rm H} \ \leq \ 10^{-4.77}$ cm$^{-3}$. Similarly,
the FOS upper limit on $N$(\ion{C}{4}) can be used to set tighter constraints
on the photoionized model in this case: the $N$(\ion{O}{6})/$N$(\ion{C}{4}) 
limit requires log $U >$ --1.33 and $n_{\rm H} \ \leq \ 10^{-5.13}$ cm$^{-3}$.

\subsubsection{Collisional Ionization}

We next consider whether the gas could be collisionally ionized. In \S 4.1
we derived $T \leq 5.8 \ \times \ 10^{4} \ ^{\circ}$K for the gas giving
rise to the main component of the \ion{H}{1} profile (the component at $v 
\approx$ 0 \kms ). According to the collisional ionization equilibrium
calculations of Sutherland \& Dopita\markcite{sd93} (1993), no \ion{O}{6} 
is produced by collisional ionization at this temperature, and in fact 
\ion{O}{2} is the dominant ionization stage of oxygen. If we assume that
the \ion{O}{6} and \ion{H}{1} absorption lines arise in the same gas 
with the same turbulent broadening, we obtain $T$ = 3.9 $\times \ 10^{4} 
\ ^{\circ}$K, which makes collisional ionization seem rather unlikely.
However, the absorber could be out of ionization and thermal equilibrium
or it could be a multiphase medium, possibilities which are discussed below.

\begin{figure}
\plotone{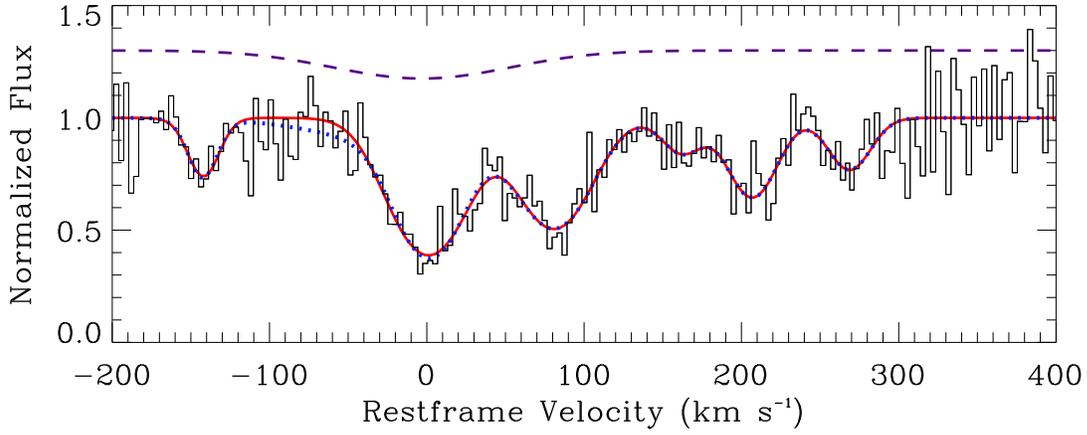}
\caption[]{Two independent fits to the \ion{H}{1} \lya profile at
\zabs\ = 0.14232 (observed spectrum is plotted with a histogram). The
solid line shows the initial best-fit with the line parameters in Table
~\ref{compprop}. The dotted line shows an alternative fit which started with
a very similar initial guess at the line parameters but which was forced
to include a broad component with a velocity and $b$-value consistent with
hot gas associated with the \ion{O}{6} absorber. This additional broad
component is shown with a dashed line offset by 0.3 units above the data.
It is very difficult to distinguish between the two fits, and we cannot
eliminate the possibility that the \ion{H}{1} profile contains a broad
component due to hot gas. Note that the STIS data have been plotted at
full (i.e., unbinned) resolution in this figure.\label{broaddemo}}
\end{figure}

Furthermore, as noted in \S 4.1, there are substantial uncertainties in the
$b$-values, and the \ion{H}{1} Doppler parameter is consistent with 
gas having $T \approx 1.6 \ \times \ 10^{5} \ ^{\circ}$K at the 3$\sigma$
level. From the \ion{O}{6} Doppler parameter we obtain $T \leq 3.5 
\ \times \ 10^{5} \ ^{\circ}$K. These temperatures are more in line with
collisionally ionized gas. Furthermore, it is entirely possible that the 
\ion{H}{1} profile contains a broad component at $v \sim$ 0 \kms\ superposed 
on a more narrow component. To demonstrate this, we show in Figure 
~\ref{broaddemo} two independent fits to the \ion{H}{1} \lya line. The
model profile indicated with a solid line is the initial best fit which has
the profile parameters summarized in Table ~\ref{compprop}. The model
profile plotted with a dotted line used a very similar initial guess at the
component parameters but was forced to include an additional component
with the same velocity as the \ion{O}{6} line and $b$ = 76 \kms , the Doppler
parameter of \ion{H}{1} in gas with $T = \ 3.5 \ \times 10^{5} \ ^{\circ}$K
and no turbulent broadening. The two profile fits are nearly indistinguishable
--- by adjusting the parameters of the other components, the profile fitting
code is able to compensate for the presence of the broad component to produce
a very similar final result. The final column density required by the profile
fitting software for the broad hot component is $N$(\ion{H}{1}) = 1.4 $\times \ 
10^{13}$ cm$^{-2}$. Unless the spectrum has very good S/N, it would
be easy to miss such a broad hot \ion{H}{1} component.  Therefore we cannot 
rule out the possibility that the gas is collisionally ionized on the basis of 
the \ion{H}{1} \lya profile properties.

Figure ~\ref{broaddemo} is not shown in order to claim that such a broad 
component is present, but rather to show that it is allowed by the data. 
However, this exercise clearly shows that if the \ion{O}{6} absorption 
occurs in hot gas with $T \approx 3 \times 10^{5} \ ^{\circ}$K, then the 
absorber must be a multiphase medium because cooler gas is required to 
account for the rest of the \ion{H}{1} \lya absorption including much of
the absorption centered at $v \approx$ 0 \kms . Given sufficient sensitivity,
metal lines such as \ion{C}{3} $\lambda$977.02 and \ion{Si}{3} 
$\lambda$1206.50 should be detectable in these cooler phases, but this may
require substantial improvement over the current signal-to-noise due to the
low \ion{H}{1} column densities.

\begin{figure}
\plotone{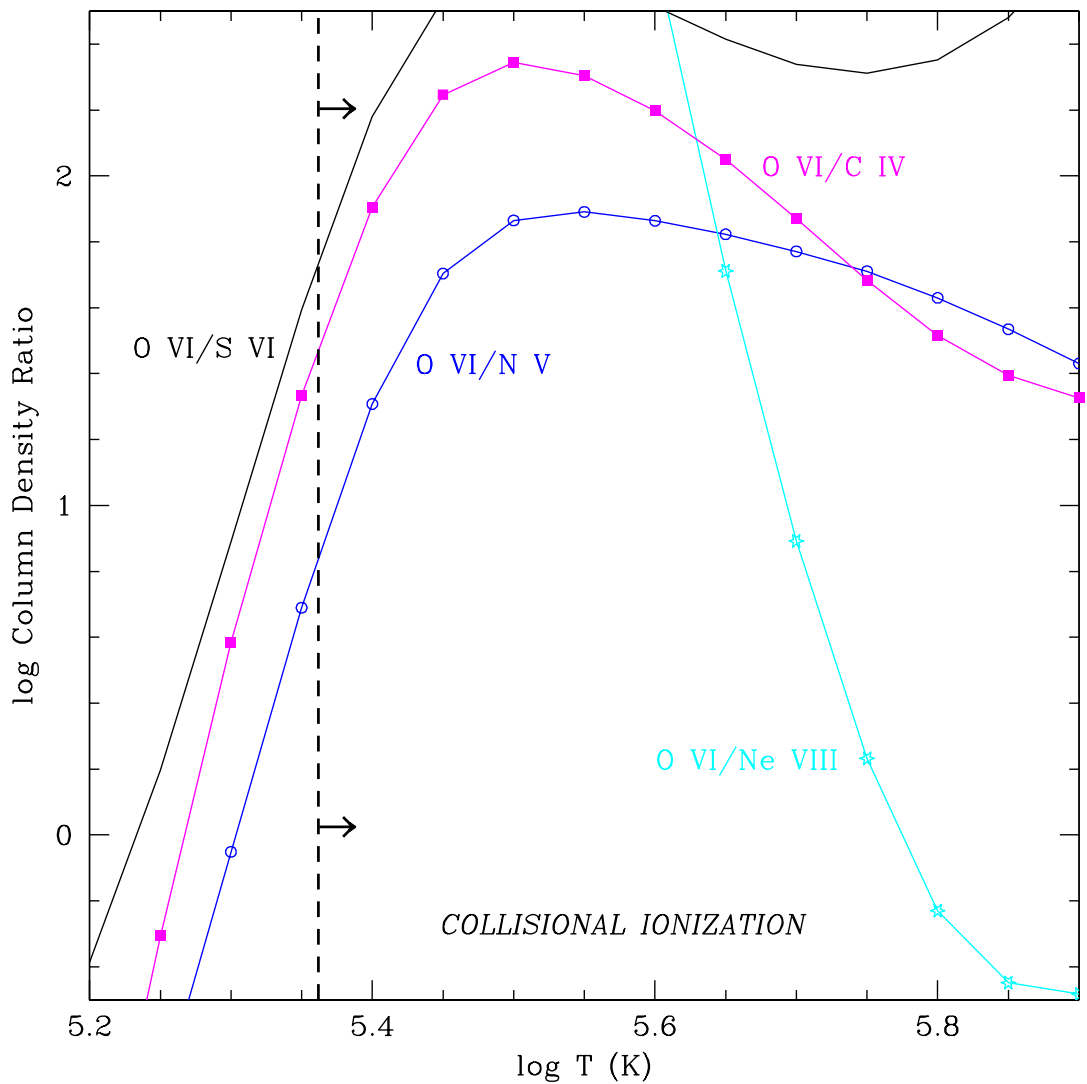}
\caption[]{Logarithmic high ion column density ratios expected in hot
collisionally ionized gas according to the equilibrium calculations of
Sutherland \& Dopita\markcite{sl93} (1993): \ion{O}{6}/\ion{N}{5} (open
circles), \ion{O}{6}/\ion{C}{4} (filled squares), and \ion{O}{6}/\ion{Ne}{8}
(open stars). The ratios are plotted versus gas temperature. The observational
lower limit on $N$(\ion{O}{6})/$N$(\ion{N}{5}) requires log $T \ \geq$ 5.36,
i.e. only the region of the plot to the right of the vertical dashed line is in
agreement with this observational constraint.\label{coll}}
\end{figure}

We next examine the high ion column density ratios predicted
for collisionally ionized hot gas. Several useful high ion column density 
ratios from the collisional ionization equilibrium calculations of 
Sutherland \& Dopita\markcite{sd93} (1993) are plotted as a function of
gas temperature in Figure ~\ref{coll}.  From this figure we see that if the
gas is in collisional ionization {\it equilibrium}, then $T$ must be greater
than $2.3 \times 10^{5} \ ^{\circ}$K to satisfy the lower limit on 
$N$(\ion{O}{6})/$N$(\ion{N}{5}), assuming [N/O] = 0. If [N/O] = --1.5, then 
$T \geq 1.8 \times 10^{5} \ ^{\circ}$K.  These are plausible temperatures 
as discussed above. 

\begin{deluxetable}{lccccc}
\footnotesize
\tablewidth{0pc}
\tablecaption{Comparison of High Ion Ratios to Non-Equilibrium Collisionally
Ionized Models\label{noneq}}
\tablehead{Column Density & Observed & Cooling & Cluster & Mixing & Conductive 
\nl
Ratio & Limit\tablenotemark{a} & Fountain\tablenotemark{b} & Cooling
Flow\tablenotemark{c} & Layer\tablenotemark{d} & Interface\tablenotemark{e}}
\startdata
\ion{O}{6}/\ion{Si}{4} & $>$13.8 & 6.2 -- 49 & 66 -- 180 & 17 -- 19 & 39 -- 210
\nl
\ion{O}{6}/\ion{C}{4}  & $>$1.6\tablenotemark{f} & 2.0 -- 10 & 5.7 -- 14 & 0.7 
-- 0.7 & 1.4 -- 4.5 \nl
\ion{O}{6}/\ion{N}{5}  & $>$5.5  & 13 -- 23  & 14 -- 21 & 7.6 -- 7.6 & 4.1 -- 7.
4 \nl
\ion{O}{6}/\ion{S}{6}\tablenotemark{g} & \nodata & 120 -- 140 & 92 -- 104 & 
\nodata & 44 -- 81 \nl
\ion{O}{6}/\ion{Ne}{8}\tablenotemark{h} & \nodata & 1.4 -- 2.4 & \nodata & \nodata & \nodata 
\enddata
\tablenotetext{a}{Based on the data in Table ~\ref{lineprop} unless otherwise
indicated.}
\tablenotetext{b}{Cooling fountain model from Benjamin \&
Shapiro\markcite{benj}
(2000), including self-photoionization, for gas initially heated to 10$^{6} \
^{\circ}$K.}
\tablenotetext{c}{Cluster cooling flow model from Edgar \&
Chevalier\markcite{edg86} (1986) for the isochoric and $T_{\rm tr} = 2.2 \times
10^{5} \
^{\circ }$K cases.}
\tablenotetext{d}{Turbulent mixing layer ratios calculated by Slavin
et al.\markcite{slav} (1993) for entrainment velocities of 25--100 \kms , a
post-mixing gas temperature of 10$^{5.0} \ ^{\circ}$K, and no dust depletion
(see their Table 4A).}
\tablenotetext{e}{Thermal conduction front model from Borkowski et
al.\markcite{bork} (1990) with an age of $10^{5.2}$ years and magnetic field
inclinations ranging from 0 to 60$^{\circ}$ relative to the front normal (see
their Figure 6). Note that in this model, these ratios increase or remain
roughly constant as the age exceeds $10^{5.2}$ years.}
\tablenotetext{f}{Based on the 4$\sigma$ upper limit on the C IV
equivalent width from Jannuzi et al. (1998), assuming the linear portion of
the curve of growth applies.  This limit is less conservative than the limits
in the third column of Table ~\ref{lineprop} because it does not account for
the uncertainty in the continuum placement.}
\tablenotetext{g}{S VI has resonance lines at 933.38 and 944.52 \AA\
which can be observed with the {\it Far Ultraviolet Spectroscopic Explorer}
(Sembach\markcite{s99} 1999). Unfortunately, at this redshift one of the S VI 
lines may blend with the Ar I 1066.66 \AA\ resonance line.}
\tablenotetext{h}{The Ne VIII resonance lines are at 770.41 and 780.32
\AA\ and therefore cannot be observed at the redshift of the O VI
absorber studied in this paper.  However, {\it FUSE} will be able to look for
this doublet in somewhat higher redshift O VI systems such as the one
studied by Savage et al.\markcite{stl98} (1998), so we include it here for
purposes of illustration (see text).}
\end{deluxetable}

However, the temperature range shown in Figure ~\ref{coll} is near the peak
of the cooling function, and gas at these temperatures is likely to cool 
faster than it can recombine, even if it has low metallicity (Sutherland \& 
Dopita\markcite{sd93} 1993), leading to an overionized condition. Consequently,
it may be more appropriate to compare the observed ratios to non-equilibrium
calculations such as cooling fountain and cluster flow models (e.g., Edgar \& 
Chevalier\markcite{edg86} 1986; Benjamin \& Shapiro\markcite{benj} 2000).
Table ~\ref{noneq} compares the observed lower limits on the 
\ion{O}{6}/\ion{Si}{4}, \ion{O}{6}/\ion{C}{4}, and \ion{O}{6}/\ion{N}{5} 
column density ratios in the \zabs\ = 0.14232 absorber to the ratios predicted 
by non-equilibrium collisionally ionized models including the cooling fountain 
calculation of Benjamin \& Shapiro\markcite{benj} (2000), the cluster cooling 
flow model of Edgar \& Chevalier\markcite{edg86} (1986), the turbulent mixing 
layer model of Slavin, Shull, \& Begelman\markcite{slav} (1993), and the 
magnetized thermal conduction interface modeled by Borkowski, Balbus, \& 
Fristrom\markcite{bork} (1990). From this table we see that the cooling 
fountain and conductive interface models are fully consistent with the current 
observational constraints on this absorber given the right choice of model 
parameters.  The turbulent mixing layer is marginally inconsistent with the 
\ion{O}{6}/\ion{C}{4} limit, but this should be confirmed with higher 
sensitivity observations of $N$(\ion{C}{4}).  However, we note that
if the turbulent mixing layer post-mixing gas temperature is increased, then
the model ironically becomes inconsistent with the \ion{O}{6}/\ion{Si}{4} 
limit as well, presumably due to increased production of \ion{Si}{4} by 
self-photoionization.

Given the available observational data, it is difficult to definitively 
constrain the ionization mechanism. It would be interesting to search for 
the \ion{Ne}{8} $\lambda \lambda$ 770.41, 780.32 doublet in \ion{O}{6} 
absorbers.  From Figures ~\ref{photo} and ~\ref{coll}, one can see that in 
the equilibrium ionization models, \ion{O}{6}/\ion{Ne}{8} $\gg$ 1 unless the 
gas has rather low density (in the photoionized case) or is relatively hot (if 
collisionally ionized).  On the other hand, $N$(\ion{O}{6}) 
$\sim N$(\ion{Ne}{8}) in the non-equilibrium cooling fountain model 
(Table ~\ref{noneq}).  Therefore detection of \ion{Ne}{8} would provide 
evidence in favor of non-equilibrium collionsional ionization.  Given 
sufficient $z$, the \ion{Ne}{8} doublet will be redshifted into the bandpass 
of the {\it Far Ultraviolet Spectroscopic Explorer} or even the STIS 
$\lambda$ range. Unfortunately, \zabs\ = 0.14232 is inadequate, and the 
\ion{Ne}{8} doublet is unobservable in this system. It is possible 
to look for the \ion{S}{6} $\lambda \lambda$ 933.38, 944.52 doublet at \zabs\ = 
0.14232 with {\it FUSE}. However, in most models this doublet is predicted to 
be substantially weaker than the \ion{O}{6} lines (see Figures ~\ref{photo} 
and ~\ref{coll} and Table ~\ref{noneq}).

\section{Discussion}

\subsection{Association with a Galaxy Group}

The fact that the \zabs\ = 0.14232 \ion{O}{6} absorption line system is 
associated with a group of galaxies strongly indicates that the absorber 
is an {\it intervening} system rather than an {\it intrinsic} absorber
which was ejected or somehow accelerated to high displacement velocity by 
the QSO. While this may seem a trivial conclusion since the \ion{O}{6} is
displaced from the QSO redshift by $\sim$23,000 \kms , during the last few
years observations of absorption variability have established that some 
highly ionized intrinsic QSO absorbers are separated from $z_{\rm QSO}$ by 
such large velocities (e.g. Hamann, Barlow, \& Junkkarinen\markcite{hbj97} 
1997) and yet are relatively narrow (i.e., not traditional broad absorption
line outflows). Therefore, it is important to find evidence that a given
\ion{O}{6} absorber is indeed intervening even if $\Delta v$ is large.
Of course, this association with a galaxy group does not necessarily indicate
that the absorption arises in the intragroup medium; it could be due to 
gas within one of the galaxies in the group. We briefly discuss some
possibilities below.

The close proximity of the \ion{O}{6} system to a galaxy group
provides a theoretical prejudice in favor of collisional ionization. The
presence of galaxies requires that intergalactic gas has collapsed in this
region of space, and simple arguments (e.g., \S 3 in Cen \& Ostriker
\markcite{co99a} 1999a) suggest that substantial shock-heating probably 
occurred as a result.  Therefore collisionally ionized gas is expected in
the vicinity (if not along the pencil-beam probed by the QSO).
It is interesting that this absorption system and associated galaxies fit the 
prediction of Mulchaey et al.\markcite{mul96} (1996) discussed in \S 1: a 
group which is possibly spiral-rich and has associated \ion{O}{6} absorption, 
as expected based on their postulated collisionally ionized intragroup medium. 
However, we really are not sure that this is a spiral-rich group (or even a 
bound group) with only four known galaxies, and it is also possible that 
the \ion{O}{6} absorption arises in the gaseous halo of a single galaxy. 
However, models of galaxy gaseous halos (e.g., Mo \& Miralda-Escud\'{e} 
\markcite{mo} 1996) also usually produce \ion{O}{6} absorption in a 
collisionally ionized hot phase. Similarly, models which produce QSO 
absorption line gas in supernova-driven winds from dwarf galaxies (e.g. 
Wang\markcite{wang} 1995) also involve substantial shock-heating and 
collisionally ionized gas.

The kinematics of the absorption also provide useful information about 
the possible origins of the system at \zabs\ = 0.14232. A fundamental question 
is whether the absorption is due to the intragroup medium or to the ISM of a 
single galaxy that happens to intercept the line-of-sight. The \ion{H}{1} \lya\ 
profile has the ``leading edge asymmetry'' that various authors (e.g., 
Lanzetta \& Bowen\markcite{lanz92} 1992) have discussed as the signature
of a moderately edge-on rotating disk. Of course, this is not a unique
interpretation of such profile asymmetry, but it provides some evidence 
in favor of the single-galaxy interpretation.  The simplicity of the 
\ion{O}{6} profile also favors this interpretation.  The velocity
dispersion of poor galaxy groups is typically one to a few hundred \kms\ 
(Zabludoff \& Mulchaey\markcite{zab} 1998), so one might expect the 
\ion{O}{6} to be spread over a larger velocity range if the absorption 
is due to the intragroup medium.  Here, though, we must recognize that the
observed \ion{O}{6} profiles are noisy and we only detect \ion{O}{6} at the
velocity of the {\it strongest} component of the \ion{H}{1} profile.  If 
the hot gas is concentrated at the center of the group, then this may be 
the only location along the line-of-sight where we have sufficient sensitivity 
to detect it, and there may be \ion{O}{6} absorption at other velocities which 
has fallen below our detection threshold.  To sort out the various 
possibilities, it would be very helpful to obtain additional STIS observations
to improve the signal-to-noise and search for \ion{O}{6} at other velocities.

\subsection{Number Density and Cosmological Mass Density}

One means to test the warm/hot gas prediction of cosmological simulations 
(\S 1) is to compare the number of \ion{O}{6} absorbers observed per unit
redshift, $dN/dz$, to the number statistically predicted from many random 
pencil-beams through the cosmological simulations.  Is the observed 
$dN/dz$ consistent with the number predicted by the cosmological models?  
We can also estimate the mean cosmological mass density traced by the 
\ion{O}{6} systems at low $z$. 

For these purposes, we combine the STIS observations
of PG0953+415 with the GHRS observations of H1821+643 from Tripp et 
al.\markcite{tls98} (1998).\footnote{After this paper was completed, Tripp, 
Savage, \& Jenkins\markcite{tsj2000} (2000) carried out an analysis 
of new STIS echelle observations of H1821+643 with the E140M mode, and we 
refer the reader to that paper for the O VI number density and 
cosmological mass density derived from an independent data set.}  We emphasize 
that the manner in which these QSOs
were selected should not bias the sample to enhance the number of 
\ion{O}{6} systems detected compared to sampling many random directions. Both 
QSOs were originally observed to study the relationship between weak \lya 
clouds and galaxies\footnote{$HST$ program IDs 6155 and 7747, see Tripp et 
al.\markcite{tls98} (1998) for details.} at $z <$ 0.3. These two objects were 
selected for this project simply because they are among the brightest known 
QSOs with $z >$ 0.2, criteria which were required to substantially increase
the sample of {\it weak} \lya absorbers with a minimal amount of {\it HST} 
time.  No consideration was given to factors which might indicate an 
enhanced likelihood of detecting the \ion{O}{6} doublet when the targets 
were selected.

To estimate $dN/dz$ of the \ion{O}{6} systems, we define a sample of 
\ion{O}{6} lines with $W_{\rm r} >$ 60 m\AA\ for both lines of the 
doublet,\footnote{The 3$\sigma$ detection limit throughout the region 
where O VI absorbers can be detected is 60 m\AA\ or better in the 
spectrum of PG0953+415 and 50 m\AA\ or better in the spectrum of H1821+643.} 
and set the maximum absorber redshift, $z_{\rm max}$, for each sight line 
to exclude any absorbers within $\mid \Delta v \mid \ \leq$ 5000 \kms\ of 
the QSO redshift and thereby avoid contaminating the sample with 
associated/intrinsic absorbers which are close to the QSO. One might argue 
that 5000 \kms\ is insufficient (see above). However, we find that the two 
\ion{O}{6} systems in the final sample are associated with galaxy groups, 
which suggests that these are indeed intervening absorbers. The lower redshift 
cutoff for each sight line, $z_{\rm min}$, was determined by the lowest 
wavelength in the observed spectrum (with a small buffer to ensure that a line 
would be recognized if at that $\lambda$) in the case of H1821+643 and by the 
wavelength at which the S/N is unacceptably low in the case of PG0953+415. 
These criteria resulted in a sample of two \ion{O}{6} absorbers within a total 
redshift path of $\Delta z$ = 0.100 (after a significant correction of 0.067 
for regions of the spectra in which we cannot detect either of the \ion{O}{6} 
lines\footnote{Both of the O VI lines were required to fall in 
unblocked regions of the spectra so that the doublet could be securely 
identified.} because they would be blocked by strong ISM or extragalactic 
lines from other redshift systems). Therefore, $dN/dz \sim$ 20 for \ion{O}{6} 
systems with $W_{\rm r} \geq$ 60 m\AA\ at $z <$ 0.3. Using the confidence 
limits from Gehrels\markcite{geh} (1986) for a sample of two absorbers, we 
derive 4 $< dN/dz <$ 63 for these \ion{O}{6} absorbers at the 90\% confidence 
level.  

For comparison, Tripp et al.\markcite{tls98} (1998) find $dN/dz = 102 
\pm 16$ for \ion{H}{1} \lya lines with $W_{\rm r} \geq$ 50 m\AA\ at $z <$ 0.3, 
and Weymann et al.\markcite{wey} (1998) report $dN/dz = 33 \pm 4$ for \lya 
absorbers at $z$ = 0 with $W_{\rm r} \geq$ 240 m\AA\ based on a large sample 
at $z <$ 1.5. In the case of low to moderate redshift \ion{Mg}{2} absorbers,
$dN/dz = 0.97 \pm 0.10$ for $W_{\rm r} \geq$ 300 m\AA\ (Steidel \& Sargent
\markcite{ss92} 1992) and $dN/dz = 2.65 \pm 0.15$ for $W_{\rm r} \geq$ 
20 m\AA\ (Churchill et al.\markcite{crcv99} 1999; see also \S 6 in Tripp, Lu, 
\& Savage\markcite{tls97} 1997). Evidently, these weak low $z$ \ion{O}{6} 
systems have a substantially larger cross section and/or covering factor than 
the \ion{Mg}{2} absorbers. Similarly, the {\it stronger} \ion{O}{6} 
absorbers at higher redshifts are less common: 
Burles \& Tytler\markcite{bt96} (1996) report $dN/dz = 1.0\pm 0.6$ for
\ion{O}{6} systems with $W_{\rm r} \geq$ 210 m\AA\ at $<z_{\rm abs}>$ = 0.9, 
and similar results (with smaller uncertainties) are derived from the larger 
sample provided by the FOS Quasar Absorption Line Key Project (B. Jannuzi 
\& R. Weymann 2000, private communication). This comparison with the \ion{O}{6}
$dN/dz$ derived from FOS data must be interpreted carefully, however, because
the FOS samples are dominated by lines which are substantially stronger and at
substantially higher redshifts than the \ion{O}{6} absorbers discussed in 
this paper.  With a larger sample of low $z$ weak \ion{O}{6} 
absorbers and smaller uncertainties in their number density,
comparison of $dN/dz$ to the space density of objects such as dwarf galaxies
may provide insight on the nature of the \ion{O}{6} systems.

Next we estimate the baryonic content of the \ion{O}{6} absorbers, expressed as
the cosmological mass density $\Omega _{b}$(\ion{O}{6}), following previous 
analogous calculations for damped \lya\ systems (e.g., Lanzetta et 
al.\markcite{lanz91} 1991) as well as \ion{O}{6} absorbers (Burles \& 
Tytler\markcite{bt96} 1996).  To estimate the density of baryons in the gaseous
component of the universe traced by \ion{O}{6}, we require information about
the metallicity of the gas and the \ion{O}{6} ionization fraction, 
$f$(\ion{O}{6}) = \ion{O}{6}/O$_{\rm total}$. In collisional ionization 
equilibrium, 
$f$(\ion{O}{6}) peaks at $\sim$0.2 (Sutherland \& Dopita\markcite{sd93} 1993), 
and similarly low peak fractions are predicted by non-equilibrium collisional 
models (e.g., Shapiro \& Moore\markcite{shape76} 1976; Benjamin \& 
Shapiro\markcite{benj} 2000). The \ion{O}{6} ion 
fraction is not much larger at peak value in photoionized gas (see appendix). 
Therefore we 
will adopt $f$(\ion{O}{6}) $\sim$ 0.2 to set a lower limit on the baryonic  
content of the \ion{O}{6} absorbers, and the following calculation is 
relatively independent of how the gas is ionized or whether or not it is 
due to the intragroup medium. For the mean cosmic 
metallicity of the \ion{O}{6} absorbers, it is less clear what value to adopt, 
but to set a lower limit on $\Omega _{b}$(\ion{O}{6}) we should set (O/H) to
a high but plausible value.  As noted in \S 4.2.1, Davis et al.\markcite{dmm99} 
(1999) and Hwang et al.\markcite{hwa} (1999) have derived metallicities of 
1/3--1/2 solar for intragroup gas in several X-ray bright groups.  Therefore 
we will initially use 1/2 solar metallicity for the calculation of 
$\Omega _{b}$(\ion{O}{6}) and then discuss how it scales with (O/H).  The mean 
cosmological mass density in the \ion{O}{6} absorbers, in units of the current 
critical density $\rho _{c}$, can be estimated as
\begin{equation}
\Omega _{b}({\rm O \ VI}) = \frac{\mu m_{\rm H} H_{0}}{\rho _{c} c 
f({\rm O \ VI})} \left( \frac{\rm O}{\rm H} \right)^{-1}_{\rm O \ VI} 
\frac{\sum_{i} N_{i}({\rm O \ VI})}{\sum_{i} \Delta X_{i}}
\end{equation}
where $\mu$ is the mean atomic weight (taken to be 1.3), 
(O/H)$_{\rm O \ VI}$ is the assumed mean oxygen abundance by number in the 
\ion{O}{6} absorption systems, $m_{\rm H}$ is the mass of hydrogen, 
$N_{i}$(\ion{O}{6}) is the total \ion{O}{6} column density and 
$\Delta X_{i}$ is the absorption distance interval (Bahcall \& 
Peebles\markcite{bah69} 1969) probed to the $i$th QSO,
\begin{equation}
\Delta X_{i} = \case{1}{2} \{ [(1 + z_{\rm max})^{2} - 1] -
[(1 + z_{\rm min})^{2} - 1]\}
\end{equation}
assuming $q_{0}$ = 0.\footnote{Over the redshift range probed by the sight 
lines to PG0953+415 and H1821+643, results are insensitive to the value 
assumed for $q_{0}$.} As in the calculation of $dN/dz$, we correct $\Delta
X_{i}$ for spectral regions blocked by strong lines. Combining the PG0953+415
STIS data and the GHRS observations of H1821+643 from Tripp et
al.\markcite{tls98} (1998) with $z_{\rm min}$ and $z_{\rm max}$ set to the
same values used for the derivation of $dN/dz$, we obtain
$\Omega _{b}({\rm O \ VI}) \gtrsim 0.0006 h_{75} ^{-1}$ assuming the mean O 
abundance is 1/2 solar.  This is a lower limit not only because 
$f$(\ion{O}{6}) and (O/H) were set to their approximate upper limits, but 
also because we have applied an equivalent width cutoff to define the sample; 
if \ion{O}{6} absorbers with $W_{\rm r} \leq$ 60 m\AA\ significantly increase  
$\sum_{i} N_{i}$(\ion{O}{6}), then the true $\Omega _{b}$(\ion{O}{6}) will be 
higher.  Note that $\Omega _{b}({\rm O \ VI})$ is inversely proportional to 
(O/H).  Decreasing the metallicity to 1/10 solar, for example, increases the 
lower limit on the \ion{O}{6} absorber baryon content to 
$\Omega _{b}({\rm O \ VI}) \gtrsim 0.003 h_{75} ^{-1}$.

To demonstrate the level of uncertainty in the cosmological mass density 
estimate due to small number statistics, we can recalculate 
$\Omega _{b}$(\ion{O}{6}) using an alternative expression analogous to 
equation (9) from Rao \& Turnshek\markcite{rt2000} (2000),
\begin{equation}
\Omega _{b}({\rm O \ VI}) = \frac{\mu m_{\rm H} H_{0}}{\rho _{c} c
f({\rm O \ VI})} \left( \frac{\rm O}{\rm H} \right)^{-1}_{\rm O \ VI}
\left(\frac{dN}{dz}\right) \frac{<N_{\rm O \ VI}>}{(1+z)}
\end{equation}
where $<N_{\rm O \ VI}>$ is the mean \ion{O}{6} column density of the 
absorption systems in the sample and again we have assumed $q_{0}$ = 0.
The advantage of this alternative expression for $\Omega _{b}$(\ion{O}{6})
is that we can employ the Gehrels\markcite{geh} (1986) small sample 
statistics to estimate the uncertainty in $dN/dz$ and then 
propagate the uncertainty into the estimate of $\Omega _{b}$(\ion{O}{6}).
With $dN/dz$ = $20^{+26}_{-13}$ and the other 
parameters set to the values assumed above, we find for the 1/10 solar 
metallicity case $\Omega _{b}({\rm O \ VI}) \gtrsim 0.003^{+0.004}_{-0.002}
\ h_{75} ^{-1}$ (error bars are $1\sigma$ uncertainties).

Bearing in mind that there is still considerable uncertainty in the lower 
limit on $\Omega _{b}$(\ion{O}{6}) due to the small sample, small redshift
path, and uncertain mean metallicity, this preliminary estimate suggests that 
the \ion{O}{6} absorbers may indeed harbor a significant fraction of the 
baryons in the universe at low $z$.  The lower limit assuming (O/H) = 1/10 
solar is comparable to the cosmological mass density of stars, 
\ion{H}{1}, and X-ray emitting galaxy group and cluster gas at low redshift 
(Fukugita, Hogan, \& Peebles\markcite{fhp98} 1998), for example.

\section{Summary}

The paper is summarized as follows.

(1) We have observed the low redshift QSO PG0953+415 with the E140M mode of
STIS, and this has revealed an \ion{O}{6} absorption line system at 
\zabs\ = 0.14232. This \ion{O}{6} absorber is highly ionized:
multicomponent \ion{H}{1} absorption is also detected
at this redshift, but no other species are detected in the STIS spectrum
including the \ion{N}{5} doublet, \ion{Si}{3}, and \ion{C}{2}.

(2) We have measured galaxy redshifts in the field of the QSO using 
the WIYN telescope, and there are at least four galaxies within $\sim$130 
\kms\ of the \ion{O}{6} absorber with projected distances ranging from 
395 kpc to 3.0 Mpc. Two of these galaxies appear to be spiral galaxies.

(3) A review of the observational constraints shows that we cannot 
definitively assert that the gas is collisionally ionized or photoionized,
but photoionization requires an uncomfortably high metallicity which is
inconsistent with theoretical expectations. If the gas is collisionally
ionized, it is likely that it is not in equilibrium or that it is
a multiphase absorber. Non-equilibrium collisional ionization models are 
consistent with the observations.

(4) Combining the STIS data on PG0953+415 with the high S/N low resolution 
GHRS observations of
H1821+643 from Tripp et al.\markcite{tls98} (1998), we identify two 
intervening \ion{O}{6} systems over a redshift path $\Delta z$ of only 0.10.
This implies that $dN/dz$ for \ion{O}{6} systems with $W_{\rm r} >$ 60 m\AA\ 
and $z <$ 0.3 is $\sim$20 with a large uncertainty due to the small number
of systems so far detected. This represents a large density for metal line 
systems since $dN/dz$ for \lya absorbers with $W_{\rm r} >$ 50 m\AA\ and 
$z <$ 0.3 is $102 \pm 16$ (Tripp et al.\markcite{tls98} 1998). We stress 
that the sample should not be biased in favor of \ion{O}{6} detection. 
If further observations confirm that $dN/dz$ is as 
large as 20 for \ion{O}{6} systems, then these absorbers may be an important 
baryon reservoir at low redshift, although this depends on the metallicity of 
the gas. If the mean metallicity is 1/2 solar, then $\Omega _{b}({\rm O \ VI}) 
\gtrsim 0.0006 h_{75} ^{-1}$.  However, if the mean metallicity is 0.1 solar, 
then $\Omega _{b}({\rm O \ VI}) \gtrsim 0.003 h_{75} ^{-1}$, which is 
comparable to the baryonic content of other known constituents of the low 
$z$ universe such as galaxies and gas in galaxy clusters.

\acknowledgements

This research has made use of software developed by the STIS Instrument
Definition Team for the reduction of STIS data, and we thank the STIS 
team for allowing us to use their software.  We also benefitted from the 
use of CLOUDY, and we are indebted to Gary Ferland for sharing this program 
in which he has invested years of development effort.  Similarly, we extend 
our thanks to Ed Fitzpatrick for the use of his profile fitting code and to 
Ken Sembach for his apparent column density software.  Valuable comments  
were provided by Dave Bowen, Romeel Dav\'{e}, Bruce Draine, Ed Jenkins, 
John Mathis, and Linda Sparke.  We especially appreciate the careful review 
of the manuscript provided by Jane Charlton which clarified and improved the 
paper. We also thank Francisco Haardt for private communication of 
the Haardt \& Madau background radiation fields in a convenient digital format,
and Buell Jannuzi and Ray Weymann for private communication regarding the
number density of \ion{O}{6} systems in the FOS Quasar Absorption Line Key
Project.
Finally, this research has made use of the APS Catalog of POSS I, which is 
supported by the NSF, NASA, and the University of Minnesota. The APS 
databases can be accessed at http://aps.umn.edu/. T. M. T. acknowledges
support from NASA through grant NAG5--30110. B. D. S. acknowledges NASA support 
through grant GO--06499.02--95A from the Space Telescope Science Institute.

\appendix

\section{Maximum Ionization Fraction of \ion{O}{6}}

In \S 5.2 we adopted $f$(\ion{O}{6}) = 0.2 to set a lower limit on the 
baryonic content of the \ion{O}{6} absorbers. Intuitively, $f$(\ion{O}{6}) = 
1.0 would seem to set a more conservative lower limit on 
$\Omega _{b}$(\ion{O}{6}). However, as noted in \S 5.2, \ion{O}{6} is not
a preferred ionization stage of oxygen, and in collisionally ionized gas 
$f$(\ion{O}{6}) peaks at $\sim$0.2 in equilibrium and non-equilibrium 
calculations. 

\begin{figure}
\plotone{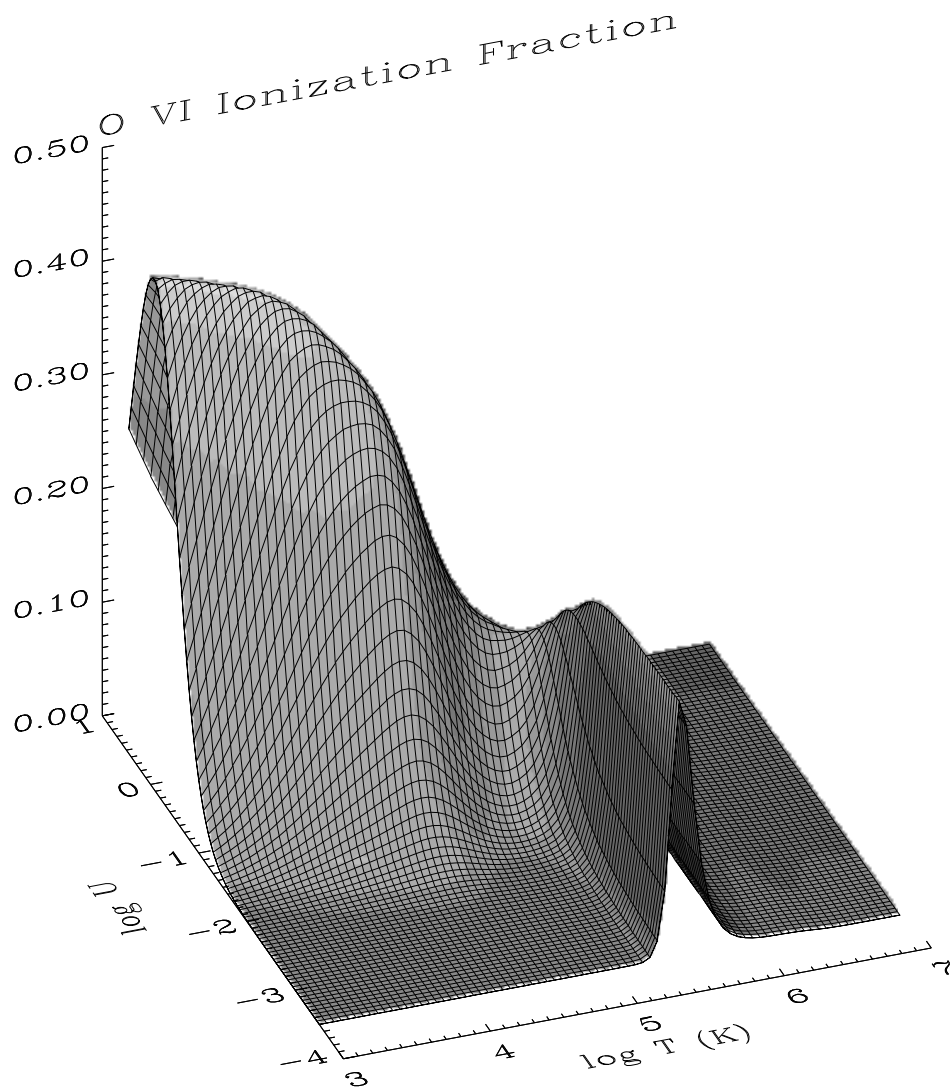}
\caption[]{Dependency of the \ion{O}{6} ionization fraction, $f$(\ion{O}{6}) =
\ion{O}{6}/O$_{\rm total}$, of a parcel of gas on the ionization parameter
$U$ and gas temperature $T$, assuming the gas is photoionized by the UV
background calculated by Haardt \& Madau\markcite{hm96} (1996) at $z \approx$
0.12 (see \S 4.2.1). It is important to note that $U$ and $T$ were held
at fixed values for the calculation of $f$(\ion{O}{6}) in each cell of this
grid to illustrate how the \ion{O}{6} ionization fraction varies with these
parameters.  However, the low temperature region of this plot is not physically
realistic because the photoionization which ionizes oxygen to this
degree will also heat the gas to log $T \ \gtrsim$ 4.\label{o6frc}}
\end{figure}

To evaluate the maximum \ion{O}{6} ionization fraction which can be 
attained if the \ion{O}{6} is created by photoionization, we have calculated
with CLOUDY an \ion{O}{6} ionization fraction grid as a function of the 
temperature and ionization parameter of a parcel of gas assuming the Haardt \& 
Madau\markcite{hm96} (1996) UV background radiation at $z$ = 0.12.
To show how $f$(\ion{O}{6}) depends on $T$ and $U$, we operated
CLOUDY in a different mode than that employed in \S 4.2.1: instead of allowing
CLOUDY to determine the 
gas temperature given the various heating and cooling processes
which are important (as was done in \S 4.2.1), we fixed the temperature and
ionization parameter at particular values in each grid cell and calculated the 
ensuing $f$(\ion{O}{6}) for a grid with 3.0 $\leq$ log $T \leq$ 7.0 and 
--3.5 $\leq$ log $U  \leq$ 0.5.  The resulting grid is shown in 
Figure ~\ref{o6frc}.

As expected based on previous calculations, when the ionization parameter is 
low enough so that photoionization is unimportant and \ion{O}{6} only has a 
significant ionization fraction in gas which is hot enough to be 
collisionally ionized, Figure ~\ref{o6frc} shows that $f$(\ion{O}{6})
has a maximum value of $\sim$0.2 at log $T$ = 5.5.  On the other hand, when the 
ionization parameter is high and photoionization dominates, 
Figure ~\ref{o6frc} indicates that 
$f$(\ion{O}{6}) could be somewhat larger {\it if the gas is very cool}. 
However, this is rather unlikely to occur because the photoionization by the 
UV background which creates the \ion{O}{6} will also heat the gas to log 
$T \gtrsim$ 4. For example, the CLOUDY model in Figure ~\ref{photo} which 
satisfies the \ion{O}{6} and \ion{N}{5} constraints requires log $T \gtrsim$ 
4.5 (see \S 4.2.1), and with this temperature constraint the maximum 
$f$(\ion{O}{6}) is 0.28.  Therefore $f$(\ion{O}{6}) = 0.2 is a reasonable 
value for placing a lower limit on the baryonic content of \ion{O}{6} 
absorbers regardless of whether the gas is photoionized or collisionally 
ionized.

\end{document}